\documentclass[11pt,aps,prd,nofootinbib,superscriptaddress, onecolumn,preprintnumbers,balancelastpage]{revtex4}
\pdfoutput=1
\usepackage{amssymb,amsmath,latexsym,graphics, graphicx,epsfig,multirow,comment,appendix, slashed, verbatim} 
\usepackage{graphicx}
\usepackage{epstopdf}
\usepackage{url}
\usepackage{appendix}
\usepackage{color}

\newdimen\tableauside\tableauside=1.0ex
\newdimen\tableaurule\tableaurule=0.4pt
\newdimen\tableaustep
\def\phantomhrule#1{\hbox{\vbox to0pt{\hrule height\tableaurule width#1\vss}}}
\def\phantomvrule#1{\vbox{\hbox to0pt{\vrule width\tableaurule height#1\hss}}}
\def\sqr{\vbox{%
  \phantomhrule\tableaustep
  \hbox{\phantomvrule\tableaustep\kern\tableaustep\phantomvrule\tableaustep}%
  \hbox{\vbox{\phantomhrule\tableauside}\kern-\tableaurule}}}
\def\squares#1{\hbox{\count0=#1\noindent\loop\sqr
  \advance\count0 by-1 \ifnum\count0>0\repeat}}
\def\tableau#1{\vcenter{\offinterlineskip
  \tableaustep=\tableauside\advance\tableaustep by-\tableaurule
  \kern\normallineskip\hbox
    {\kern\normallineskip\vbox
      {\gettableau#1 0 }%
     \kern\normallineskip\kern\tableaurule}%
  \kern\normallineskip\kern\tableaurule}}
\def\gettableau#1 {\ifnum#1=0\let\next=\null\else
  \squares{#1}\let\next=\gettableau\fi\next}

\newcommand{\gsim}{\lower.7ex\hbox{$\;\stackrel{\textstyle>}{\sim}\;$}}
\newcommand{\lsim}{\lower.7ex\hbox{$\;\stackrel{\textstyle<}{\sim}\;$}}

\def\eg{{\it e.g.}}

\newcommand{\GeV}{\,\mathrm{GeV}}

\newcommand{\be}{\begin{eqnarray}}
\newcommand{\ee}{\end{eqnarray}}
\newcommand{\bea}{\begin{eqnarray}}
\newcommand{\eea}{\end{eqnarray}}

\newcommand{\bef}{\begin{figure}[htbp]\begin{center}}
\newcommand{\eef}{\end{center}\end{figure}}

\newcommand{\GG}{\gamma \gamma}

\newcommand{\sigmaWW}{\sigma_{_{WW}}}
\newcommand{\sigmaZZ}{\sigma_{_{ZZ}}}
\newcommand{\sigmaGG}{\sigma_{\gamma \gamma}}
\newcommand{\sigmaGZ}{\sigma_{\gamma_Z}}
\newcommand{\sigmaann}{\sigma_\mr{ann}}

\newcommand{\NGG}{N_{\gamma \gamma}}
\newcommand{\NGZ}{N_{\gamma_Z}}
\newcommand{\Nann}{N_\mr{ann}}
\newcommand{\thetaGZ}{\theta_{\gamma_Z/\gamma\gamma}}
\newcommand{\cms}{\mbox{ cm}^3/\mbox{s}}

\newcommand{\mr}[1]{\mathrm{#1}}

\newcommand{\eref}[1]{Eq.~(\ref{#1})}

\begin{document}
\begin{flushright}
\mbox{\normalsize SLAC-PUB-15141}
\end{flushright}
\vskip 80 pt

\title{

Illuminating the 130 GeV Gamma Line with Continuum Photons }

\author{Timothy Cohen}
\affiliation{
SLAC, Stanford University, Menlo Park, CA 94025}

\author{Mariangela Lisanti}
\affiliation{
Princeton Center for Theoretical Science, Princeton University, Princeton, NJ 08544}

\author{Tracy R. Slatyer}
\affiliation{
School of Natural Sciences, Institute for Advanced Study, Princeton, NJ 08540}

\author{Jay G. Wacker}
\affiliation{
SLAC, Stanford University, Menlo Park, CA 94025}

\begin{abstract}
\vskip 15 pt
\begin{center}
{\bf Abstract}
\end{center}
\vskip -8 pt
$\quad$ 
There is evidence for a 130 GeV $\gamma$-ray line  at the Galactic Center in the \emph{Fermi} Large Area Telescope data.  Dark matter candidates that explain this feature should also annihilate to Standard Model particles, resulting in a continuous spectrum of photons.  To study this continuum, we analyze the \emph{Fermi} data down to 5 GeV, restricted to the inner $3^\circ$ of the Galaxy.  We place a strong bound on the ratio of continuum photons to monochromatic line photons that is independent of uncertainties in the dark matter density profile.  The derived constraints exclude neutralino dark matter as an explanation for the line.
\end{abstract}

\maketitle
\newpage

%%%%%%%%%%%
\section{Introduction}
%%%%%%%%%%%

Astrophysical searches for dark matter are a critical component of the experimental effort to explore the dark sector (\cite{Bergstrom:2012fi, Jungman:1995df}).  The strategy is to look for observable products of dark matter annihilation, both in the Milky Way halo and beyond.  A variety of experiments are currently underway (see \cite{Carr:2006cw} for a review), exploring many different annihilation products over a broad range of energies and target spatial regions.  A particularly important potential astrophysical signal is a monochromatic $\gamma$-ray line, with energy corresponding to the mass of the annihilating dark matter.  The discovery of such a line would be a ``smoking gun'' signature of dark matter because no astrophysical backgrounds are known to generate a peak in the $\gamma$ spectrum. 

The \emph{Fermi} Large Area Telescope (LAT) is currently taking data, and continually improving its sensitivity to features in the $\gamma$ spectrum that could be signatures of dark matter~\cite{Ackermann:2012qk, Ackermann:2012nb, Ackermann:2011wa, Abdo:2010nc, Abdo:2010ex, Cotta:2011pm}.  A recent analysis of the \emph{Fermi} $\gamma$ spectrum from 20--200 GeV has found preliminary evidence for a sharp feature around the Galactic Center corresponding to $E_\gamma \simeq 130\GeV$~\cite{Weniger:2012tx,Bringmann:2012vr}.  The analysis by~\cite{Weniger:2012tx} finds 3.3 $\sigma$ evidence (including the look-elsewhere effect) for such a line in a region of the sky that extends roughly 15$^{\circ}$ above and below the Galactic plane.  Based on an observation of $\sim 50$ photons, the tentative signal corresponds to dark matter with best fit mass $m_{\chi} = 129.8$ GeV and annihilation cross section $\sigmaGG v\simeq 1.27\times10^{-27}\cms$, if the dark matter density follows an Einasto profile.  

Following~\cite{Weniger:2012tx}, the authors of \cite{Tempel:2012ey, Su:2012ft} have confirmed the presence of the 130 GeV line and have strengthened the case that this excess could be due to dark matter annihilations.  The authors showed that the signal is concentrated in a $\sim 3^{\circ}$ radius region about the Galactic Center.  Additionally, they obtain better fits if the chosen region is off-set from the center of the galaxy by $\mathcal{O}(1^\circ)$, and if the signal consists of a pair of lines at 111 and 129 GeV~\cite{Su:2012ft}. 

The origin of the 130 GeV feature in the \emph{Fermi} spectrum remains a mystery.  At present, the official search by the \emph{Fermi} collaboration has found no evidence for monochromatic $\gamma$ lines~\cite{Ackermann:2012qk}.  For a 130 GeV dark matter mass, this analysis sets an upper limit of 1.0$\times10^{-27}\cms$ for an Einasto profile, which is in tension with the purported signal.  However, the two analyses are fundamentally different in their approaches.  The \emph{Fermi} collaboration searched for lines using the all-sky $\gamma$-ray maps; the analysis was done for $|b| > 10^{\circ}$ plus a $20^{\circ}\times20^{\circ}$ square at the Galactic Center using the \texttt{Pass 6} data.  In contrast, the search regions in~\cite{Weniger:2012tx} were defined to optimize the significance of a dark matter signal and the analysis was performed using the \texttt{Pass 7} data.

Our goal here is to explore constraints on the properties of dark matter that can be obtained from the \emph{Fermi} data under the assumption that the 130 GeV line is due to annihilating dark matter.  In a large class of weakly interacting dark matter models, one expects the annihilation of dark matter into a pair of photons to arise from loops of states that carry electroweak charges.  One consequence is that there is a non-zero annihilation rate into $\gamma Z^0$ and/or $\gamma h$.  If the particles in the loop are lighter than the dark matter, annihilation into these states can dominate and a continuum spectrum of photons results from the subsequent decay of the annihilation products.  This is the case for the neutralino of the minimal supersymmetric Standard Model (MSSM)~\cite{Jungman:1995df}, as well as non-supersymmetric extensions of the Standard Model, such as Universal Extra Dimensions \cite{Bergstrom:2004nr,Bertone:2009cb}, Little Higgs models \cite{BirkedalHansen:2003mpa,Perelstein:2006bq}, and more generic weakly interacting (WIMP) dark matter models \cite{Cirelli:2005uq,Cirelli:2007xd,Cirelli:2008id,Cohen:2011ec}.

In this work, we constrain the ratio of the number of continuum photons to the number of photons responsible for explaining the 130 GeV $\gamma$ line.  The expression for the photon flux is given by
\be
\Phi(E, \theta) = \frac{1}{2} \frac{\sigmaann v}{4\,\pi} \sum_f \frac{\mr{d} n_{\gamma}^f}{\mr{d} E} \mr{Br}_f \int_\mr{LOS} \mr{d} \ell(\theta) \frac{\rho(\ell)^2}{m_\chi^2},
\ee
where $\sigmaann$ is the total dark matter annihilation cross section into final states $f$, $\mr{d} n_{\gamma}^f/\mr{d} E$ is the differential photon spectrum resulting from each $f$, $\mr{Br}_f$ is the branching fraction into $f$, $\ell(\theta)$ is the line of sight (LOS) as measured at an angle $\theta$ from the Galactic Center, and $\rho(\ell)$ is the dark matter density along the LOS.  Constraints on the ratio of photons from different final states are particularly powerful because dependencies on the dark matter density profile cancel out.  As a result, the constraints derived in this paper are independent of astrophysical uncertainties.      

We present two versions of the constraint on the dark matter continuum contribution.  The first requires that the continuum photons not supersaturate the observed \emph{Fermi} data.  The second involves a log likelihood fit of the shape of the dark matter and background spectra to the \emph{Fermi} data.  The second constraint is much stronger, but relies on the assumption that the astrophysical background is well-described by a single power-law.  Our derived constraints exclude the neutralino explanation for the 130 GeV $\gamma$ line.

We begin in Sec.~\ref{sec:130GeVGammaLine} by discussing the \emph{Fermi} data used in this analysis and the details of our fitting procedure.  In Sec.~\ref{sec:ContConstraint}, we provide the constraints on the ratio of continuum photons from annihilations to $W^+W^-$ and $Z^0Z^0$ to the number of photons that yield the $\gamma$ line.  In Sec.~\ref{sec:MSSM}, these constraints are applied to neutralino dark matter.  We summarize our conclusions in Sec.~\ref{sec:Discussion}.  There are three appendices to the article that provide the counts per bin extracted from the public \emph{Fermi} data used in this analysis (Appendix~\ref{AppA}),  discuss our computation of the \emph{Fermi} Instrument Response Function for energy dispersion (Appendix~\ref{AppB}), and provide constraints for dark matter annihilations to the additional final states $b\overline{b}$, $\tau^+\tau^-$, and $\mu^+\mu^-$ (Appendix~\ref{AppC}).

%%%%%%%%%%%%%%
\section{A 130 GeV Gamma Line}
\label{sec:130GeVGammaLine}
In this section, we describe the \emph{Fermi} data used in our analysis and the methodology employed in our log likelihood fitting procedure.  As an application of our statistical procedure, we consider the best fit for a pair of lines, which is expected if the dark matter annihilates to both $\gamma \gamma$ and $\gamma Z^0$.

%%%%%%%%%%%%%%
\subsection{The \emph{Fermi} Data}
%%%%%%%%%%%%%%

The \emph{Fermi} LAT 3.7 year public data that is used in this analysis includes the direction of arrival and the reconstructed energy for each measured photon in the \texttt{Pass 7\_Version 6} release.\footnote{http://heasarc.gsfc.nasa.gov/FTP/fermi/data/lat/weekly/p7v6}  A standard zenith angle cut is applied to exclude events at angles greater than $100^\circ$.  The livetime/exposure is computed as recommended for diffuse analyses:\footnote{http://fermi.gsfc.nasa.gov/ssc/data/analysis/documentation/Cicerone/Cicerone\_Likelihood/Exposure.html} specifically, we use option 2 with the criteria specified in Tab.~\ref{Tab: cuts}.  
%$z_\mr{max} = 100$ and $\mr{filter} = (\mr{DATA}\_\mr{QUAL} == 1\, \&\&\, \mr{LAT}\_\mr{CONFIG} == 1\, \&\& \,\mr{ABS(ROCK}\_\mr{ANGLE)} < 52)$.  
The $\mr{DATA}\_\mr{QUAL}$ filter excludes time periods when the quality of the data is compromised, \eg~solar flares, the $\mr{LAT}\_\mr{CONFIG}$ flag includes data from when the LAT is in nominal science configuration, and the $\mr{ABS(ROCK}\_\mr{ANGLE)}$ cut excludes periods when the Earth limb is in the field of view.  We use the class of events designated \texttt{ULTRACLEAN}, which have a lower effective area but also a lower background than the \texttt{SOURCE} class.

\begin{table}
\begin{tabular}{|ccc|}
\hline
$z_\mr{max}$ &=& 100\\
 $\mr{DATA}\_\mr{QUAL}$ &=& 1\\
$ \mr{LAT}\_\mr{CONFIG}$& = &1\\
 $\mr{ABS(ROCK}\_\mr{ANGLE)}$ &$<$& 52\\
 \hline
\end{tabular}
\caption{\label{Tab: cuts} Criteria used to filter events.}
\end{table}

Following \cite{Su:2012ft}, we restrict our analysis to the inner $3^\circ$ radius region around the Galactic Center and neglect possible enhancements from an offset along the plane \cite{Su:2012ft}.  Unless explicitly stated, all results use data where the area within 1 degree of the Galactic Center is masked to reduce background contributions.  We restrict to the energy range 5--200 GeV to minimize uncertainties due to the point spread function (PSF).  The \emph{Fermi} LAT is designed to measure photons from around $20 \mbox{ MeV}$ to many hundred GeV.  The PSF, which encodes the uncertainty in the reconstructed position in the sky, starts to
grow rapidly below a GeV.  Specifically, the $68 \%$ containment radius of the PSF is about $0.9^\circ$ at 1 GeV and decreases with energy, approaching $\sim 0.2^\circ$ at high energies.  

Appendix \ref{sec:CountsPerBin} provides the counts per bin for the relevant region of the sky when the inner degree is both masked and unmasked.  The photon counts are given for $N_{\text{bins}}=128$ energy bins from \mbox{5.1--198 GeV}.

%%%%%%%%%%%%%%
\subsection{Fitting The Data}
\label{sec: fitsignal}
%%%%%%%%%%%%%%

For concreteness, we assume that the signal arises from a WIMP of mass $m_{\chi}$ annihilating into $\gamma\gamma$ and/or $\gamma Z^0$, thereby producing at most two lines in the photon spectrum at energies 
\begin{equation}
E_{\GG} = m_{\chi} \quad \quad \text{and} \quad \quad E_{\gamma_Z} = m_{\chi} \Bigg( 1- \frac{m_Z^2}{4 m_{\chi}^2} \Bigg).
\label{eq: signal}
\end{equation}
The WIMP may also annihilate into final states (\eg, $W^+W^-$, $Z^0\,Z^0$, $b\,\overline{b}$,  $\tau^+\tau^-$,  $\mu^+\mu^-$, etc.) whose decay products shower and hadronize to produce a continuum photon contribution.  Assuming that the background is a falling power-law parametrized by $\alpha, \beta$, the observed photon spectrum expected from this model is
\begin{equation}
\phi(E) = 
C_\mr{EA}(E)\left[\beta \Bigg( \frac{E}{\text{100 GeV}}\Bigg)^{-\alpha} 
+ \NGG D( E, E_{\gamma \gamma})
+\NGZ D( E,  E_{\gamma_Z}) + \Nann \frac{\mr{d} \overline{n}_\gamma}{\mr{d} E}\left(E,\,m_\chi\right)\right],
\end{equation}
where $\NGG$, $\NGZ$, and $\Nann$ are the normalizations of the separate signal components.  The function $D(E, E_{\text{true}})$ is the energy dispersion about the true signal energy and is derived using the \emph{Fermi} Instrument Response Function (IRF) obtained from the publicly available Science Tools\footnote{http://fermi.gsfc.nasa.gov/ssc/data/analysis/} --- see Appendix~\ref{sec:FermiIRF} for a detailed discussion.  The normalized differential distributions for different annihilation final states, denoted $\mr{d} \overline{n}_\gamma/\mr{d} E$, are obtained using \texttt{Pythia} version 8.165 \cite{Sjostrand:2000wi} to generate the spectra.  $C_\mr{EA}$ is a corrective factor that accounts for the change in effective area in the 3$^{\circ}$ region about the Galactic Center, as a function of energy.  

\begin{figure}[b!] %  figure placement: here, top, bottom, or page
   \centering
   \begin{tabular}{cc}
   \includegraphics[width=.9\textwidth]{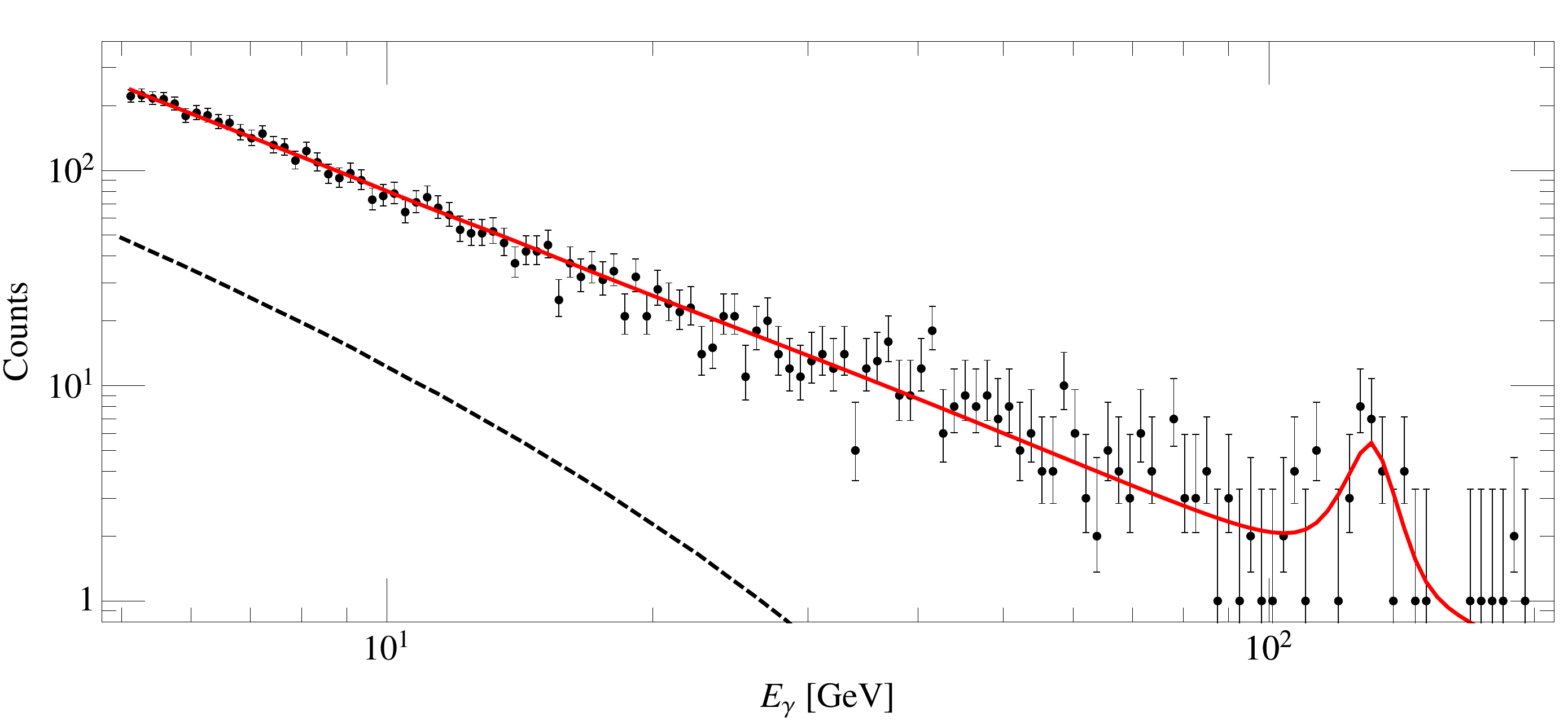}  
   \end{tabular}
   \caption{Photon counts within 3$^{\circ}$ degrees of the Galactic Center with the inner degree masked.  The solid red line shows the best fit model given in \eref{eq:BestFitModel}, assuming no continuum contribution.  The dashed black line shows the continuum spectrum for a 130 GeV dark matter annihilating into $W^+W^-$ (arbitrary normalization); the spectrum for $Z^0Z^0$ is indistinguishable.}
   \label{fig:BestFit}
\end{figure}
For Poisson-distributed data, the best fit values of the parameters $\alpha, \beta, \NGG, \NGZ$, and $\Nann$ are obtained by maximizing the likelihood function 
\begin{equation}
\ln \mathcal{L} (\alpha, \beta, \NGG, \NGZ, \Nann) = 
\sum_{k=1}^{N_{\text{bins}}} n_{k} \cdot \ln \phi_k - \phi_k - \ln n_k!,
\label{eq: loglikelihood}
\end{equation}
where $n_k$ is the observed photon count and $\phi_k = \int_{E_{\text{min}}^k}^{E_{\text{max}}^k} \phi(E) \mr{d}E$ for the $k^{\text{th}}$ bin spanning $\left[E^k_{\text{min}}, E^k_{\text{max}}\right]$.  The confidence region about the maximum likelihood, $\ln \mathcal{L}_{\text{max}}$, is determined by
\begin{equation}
\ln \mathcal{L} \geq \ln \mathcal{L}_{\text{max}} -  \Delta \ln \mathcal{L},
\end{equation}
where $2 \Delta \ln \mathcal{L} = \Delta \chi^2$ and the number of degrees of freedom (d.o.f.) is the number of fit parameters.  

Next, we use this statistical procedure to show that the photon spectrum in the region of interest is consistent with the presence of a photon line.  For now, we assume that the photon continuum does not contribute to the signal, reserving the case where $\Nann > 0$ for the next section.  Scanning over $m_{\chi}$ and 
\begin{eqnarray}
\thetaGZ  \equiv \arctan \frac{ \NGZ}{\NGG} ,
\end{eqnarray}
while maximizing over $\alpha, \beta,$ and $\NGG$, we find that the best fit point corresponds to 
\bea
\left\{m_{\chi}/\text{GeV}, \alpha, \beta, \NGG, \thetaGZ \right\}_{\text{max}} &=& \left\{130, 2.67, 0.88, 30.3, 0\right\}\quad \quad \mbox{(unmasked)};\nonumber\\
&&\label{eq:BestFitModel}\\
\left\{m_{\chi}/\text{GeV}, \alpha, \beta, \NGG, \thetaGZ \right\}_{\text{max}} &=& \left\{130, 2.62, 0.80, 31.6, 0\right\}\quad \quad \mbox{(masked)},\nonumber
\eea
where masked (unmasked) refers to removing (including) data within 1 degree of the Galactic Center.  The significance of the best fit point relative to the null model (power-law background) is 5.5 $\sigma$ for both the masked and unmasked cases, not including look-elsewhere.\footnote{The best fit null model is $\{\alpha, \beta\}_{\text{null}} = \{2.65, 0.95\}$ for the unmasked case and $\{\alpha, \beta\}_{\text{null}} = \{2.58, 0.87\}$ for the masked case.}  Masking a 1 degree radius circle around the Galactic Center has little effect on the best fit dark matter parameters, though it prefers more shallow power-law backgrounds.  From this point onwards, we will only consider the masked data.

\begin{figure}[t] %  figure placement: here, top, bottom, or page
   \centering
   \quad\includegraphics[width=.62\textwidth]{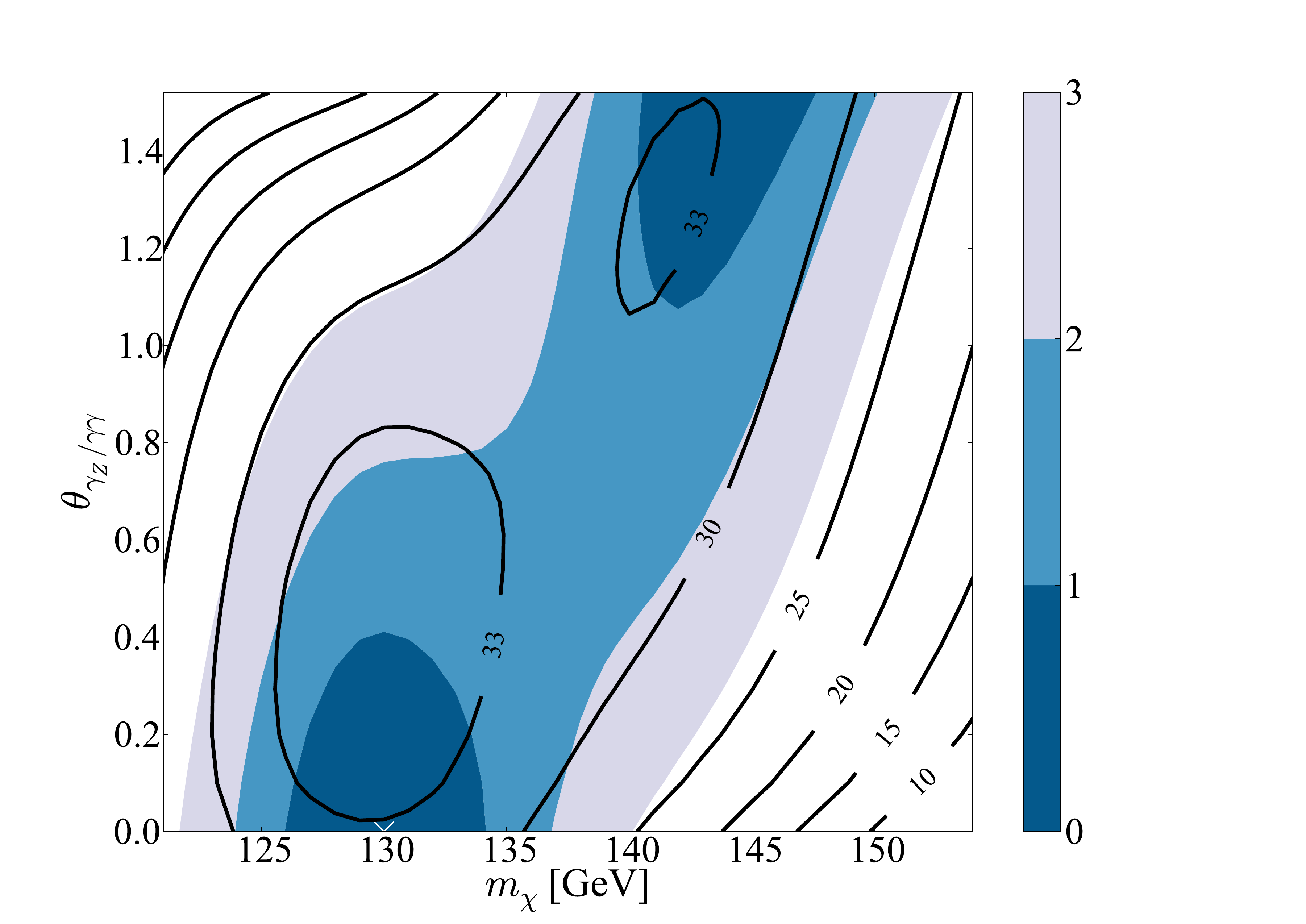} 
   \caption{Regions of 1, 2, and 3 $\sigma$ significance (filled contours) for $\thetaGZ = \arctan \NGZ/\NGG$ as a function of mass for the case $\Nann = 0$.  The 1, 2, and 3 $\sigma$ contours refer to $\Delta \ln \mathcal{L} = 1.76, 4.01,$ and $7.08$ (3 d.o.f.).  The solid lines are contours of $\NGG + \NGZ$.  The best fit point, marked with a white cross at $m_\chi = 130 \mbox{ GeV}$ and $\thetaGZ = 0$, is given in \eref{eq:BestFitModel}.  This figure was made using the masked data.  The analogous plot for the unmasked case is qualitatively the same.}
   \label{fig:FitGammaLine}
\end{figure}

Figure~\ref{fig:BestFit} shows the spectrum of photon counts in the region of interest.  The solid red line corresponds to the best fit model in \eref{eq:BestFitModel} obtained by maximizing the likelihood function over the energy range from 5--200 GeV.  The spectrum is well-characterized by a single falling power-law and a peak at 130 GeV comprised of $\sim$30 photons.  

Figure~\ref{fig:FitGammaLine} shows the 1, 2, and 3 $\sigma$ contours for points in the $\thetaGZ-m_{\chi}$ plane.  The best fit point is marked by the white ``X."  There is a clear symmetry in the significance contours, with regions about 130 and 145 GeV each within $1\,\sigma$ of the best fit model.  For the case of a 145 GeV dark matter, all the photons in the 130 GeV line are due to dark matter annihilation to $\gamma Z^0$.

Both~\cite{Su:2012ft} and~\cite{Rajaraman:2012db} note that the presence of two lines at $\sim130$ and $115$ GeV is a slightly better fit to the data; we reproduce the results in~\cite{Rajaraman:2012db} when redoing our analysis with the data from~\cite{Weniger:2012tx}. 
However, for the region of interest analyzed in this work, the data prefer a single line corresponding to a DM mass of either 130 or 145 GeV, although two lines are consistent within 1 $\sigma$.  

%%%%%%%%%%%%%%
\section{Continuum Constraint}
\label{sec:ContConstraint}
Now that we have demonstrated the presence of a line (or pair of lines) in the \emph{Fermi} data at the Galactic Center, we explore correlated photon signals that are important for a wide class of models.  If the dark matter annihilates to Standard Model particles beyond $\gamma\gamma$ and/or $\gamma Z^0$, additional photons are produced in the decay of these states.  
The \emph{Fermi} data has been used to place constraints on the resulting inclusive photon spectrum using a variety of methods \cite{Ackermann:2012qk, Ackermann:2012nb, Ackermann:2011wa, Abdo:2010nc, Abdo:2010ex, Cotta:2011pm, GeringerSameth:2011iw,GeringerSameth:2012sr}.  Currently, the strongest of these bounds comes from a search for $\gamma$ rays originating from 10 dwarf galaxies in the Milky Way over a 24 month period \cite{Ackermann:2011wa}.  For annihilations to $W$ bosons and $m_{\chi} \simeq 130 \mbox{ GeV}$ the $95 \%$ confidence bound is
\be\label{eq:DwarfGalaxyConstraint}
\sigmaWW v \lesssim 10^{-25} \mbox{ cm}^3/\mbox{s} \quad \quad \mbox{(\emph{Fermi} dwarf Galaxy constraint)}.
\ee
The constraint from stacked dwarfs rules out large regions of parameter space for models that can potentially explain the 130 GeV $\gamma$ line.  In the MSSM, for example, assuming the neutralino makes up all the dark matter, this constraint rules out all models with $\frac{1}{2}\sigmaGZ v +\sigmaGG v\gtrsim \mathcal{O}\left(10^{-28}\,\cms\right)$, except for wino-bino mixed neutralinos.

Here, we derive a constraint on the ratio of the number of continuum photons to the number of photons in the peak using the data from the Galactic Center.  Specifically, we constrain the ratio 
\be\label{eq:RTheory}
R^\mr{th}  \equiv \frac{\sigma_\mr{ann}}{2\, \sigmaGG+\sigmaGZ},
\ee
derived for a given theory,\footnote{When $R^\mr{th}$ is $\mathcal{O}(1)$, one must be careful to include the contribution to the continuum spectrum from $\chi \chi\rightarrow \gamma Z^0$, which has a slightly different shape than $\chi \chi\rightarrow Z^0 Z^0$ from kinematic effects.  We neglect this subtlety for the constraints presented in this paper.}
by comparing it with the associated quantity obtained from observation
\be\label{eq:RObserved}
R^\mr{ob} \equiv\frac{1}{n_\mr{ann}^\gamma} \frac{\Nann}{\NGG+\NGZ},
\ee
where $\sigma_\mr{ann}$ is the total dark matter annihilation cross section, and $\sigmaGG$ ($\sigmaGZ$) is the annihilation cross section to $\gamma \gamma$ ($\gamma Z^0$).  $\Nann$ is the number of photons in the continuum spectrum that results from the process that dominates $\sigma_\mr{ann}$, and $\NGG$ ($\NGZ$) are the number of photons in the peak(s) attributed to dark matter annihilations to $\gamma \gamma$ ($\gamma Z^0$).  $n_\mr{ann}^\gamma$ is the total number of photons per annihilation in the considered energy range.  We do not include this factor in the definition of $R^\mr{th}$ because it depends on the energy range of interest and will be different for the ``supersaturation" constraint (Sec.~\ref{sec:SupersaturationConstraint}) and the ``shape" constraint (Sec.~\ref{sec:ShapeConstraint}), discussed below.

In the following subsection, we constrain $R^\mr{ob}$ by requiring that the continuum photons do not supersaturate the data.  The result is model-independent, and only depends on the final state annihilation products.  Because the bound is a ratio, it is independent of any astrophysical uncertainties and applies to scenarios where the annihilating particle is a subdominant component of the dark matter \cite{Acharya:2012dz}. We then go on to present an even stronger bound using the full shape information of the continuum spectra.  Although this constraint is significantly stronger, it does depend on whether a single power-law explains the background from 5--200 GeV.  

In what follows, we focus on dark matter annihilations to $W^+W^-$ and $Z^0Z^0$.  Shape constraints for final states $b\overline{b}$, $\tau^+\tau^-$, and $\mu^+\mu^-$ are given in Appendix~\ref{sec:AlternateFinalStates}.

\subsection{Constraint from Supersaturation}\label{sec:SupersaturationConstraint}

In this subsection, we derive a constraint on $R^\mr{ob} $ as defined in \eref{eq:RObserved} that is independent of any background model assumptions.  This constraint arises from the fact that the continuum contribution should not supersaturate the data.  It is conservative in that it assumes that the entirety of the photon spectrum is due to signal, with no background contribution.  

\begin{figure}[t!] %  figure placement: here, top, bottom, or page
   \centering
   \includegraphics[width=.58\textwidth]{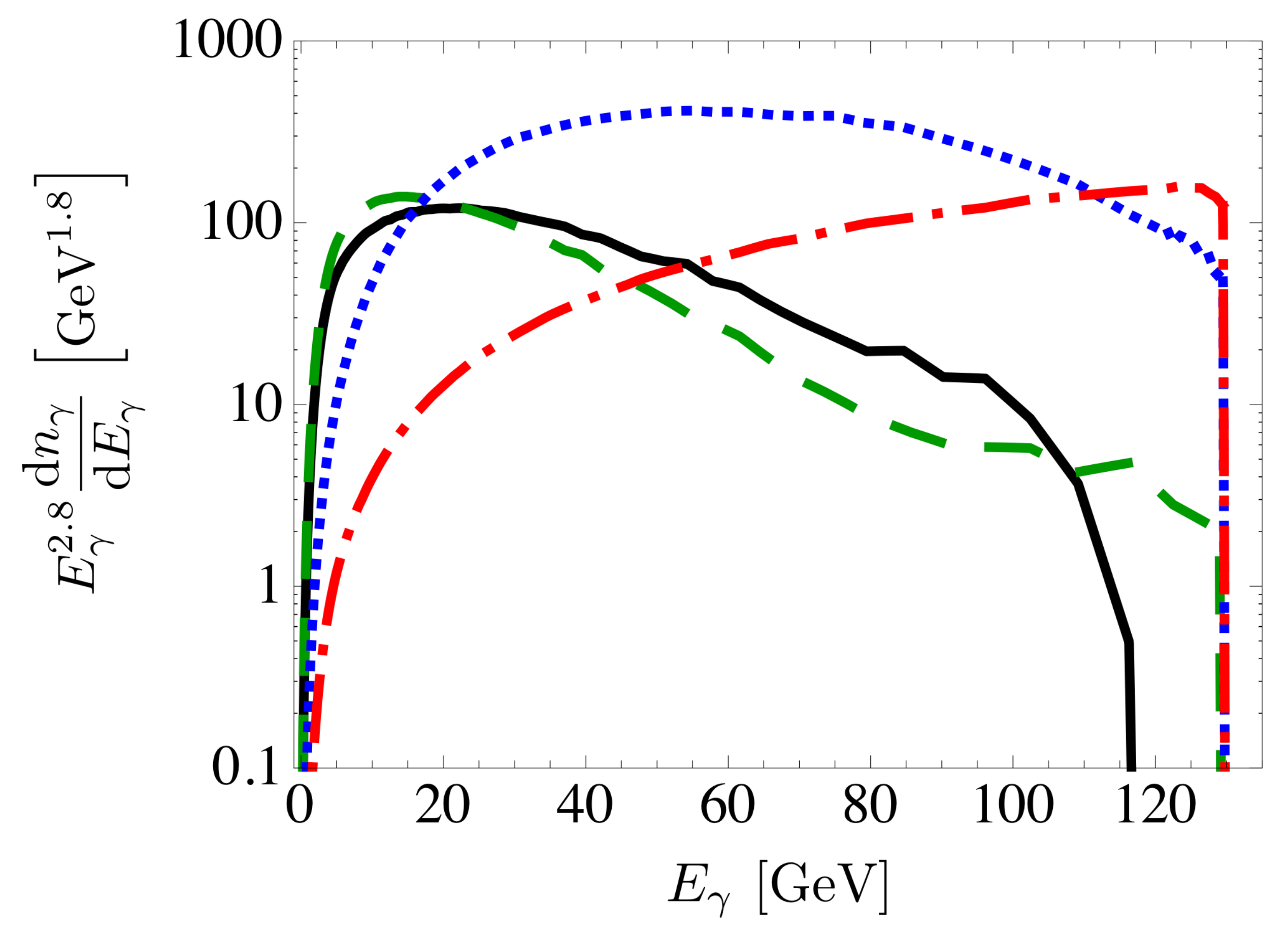}
   \vskip -10pt
   \caption{Continuum photon spectra $\times \, E_\gamma^{2.8}$ as a function of $E_\gamma$ for dark matter annihilating into $W^+W^-,\,Z^0\,Z^0$ [black, solid], $b\overline{b}$ [green, dashed], $\tau^+\tau^-$ [blue, dotted], and $\mu^+ \mu^-$ [red, dot-dashed] .  The dark matter mass is taken to be 130 GeV.  }
   \label{fig:ContPhotSpectra}
\end{figure}

To obtain the optimal supersaturation exclusion, we must select an energy bin where the number of continuum photons peaks relative to the background.  The spectral index $\alpha$ of the $\gamma$ continuum background is expected to follow that of the proton spectrum $\alpha_p$ from 5--100 GeV (see \cite{Kamae:2004xx} and references therein).  For concreteness, we use the measurement from the PAMELA collaboration in the range $E_p = 30-80 \mbox{ GeV}$:  $\alpha_p = 2.801\pm 0.007 \mbox{ (stat) } \pm 0.002 \mbox{ (syst) }$ \cite{Adriani:2011cu}.  Note that this is a local measurement and may differ at the Galactic Center.  It does, however, provide an independent determination of the spectral index and gives a reference point from which to select an optimal energy bin.

In Fig.~\ref{fig:ContPhotSpectra}, we show $E_{\gamma}^{2.8}$ multiplied by the differential continuum photon spectrum as a function of $E_{\gamma}$, where $E_{\gamma}$ is the photon energy, for $W^+W^-$, $Z^0\,Z^0$ [black, solid], $b\overline{b}$ [green, dashed],  $\tau^+\tau^-$ [blue, dotted], and $\mu^+ \mu^-$ [red, dot-dashed] final states.  The spectra are generated using \texttt{Pythia} version 8.165~\cite{Sjostrand:2000wi}.  The location of this peak is approximately at 15 GeV for $W^+W^-$, indicating where the continuum spectrum should peak over the power-law background for this final state.

The shapes of $W^+W^-$, $Z^0\,Z^0$, and $b\overline{b}$ fall off sufficiently fast as $E_\gamma \rightarrow m_\chi$.  However, for $\tau^+ \tau^-$ and $\mu^+\mu^-$, the continuum spectrum can make a non-trivial contribution at $E_\gamma \simeq m_\chi$.  For the supersaturation constraint, the number of photons in the peak must be independent of the continuum, and it is therefore important that the continuum contribution to energies near $m_\chi$ be negligible.  This is the case for $W^+W^-$, $Z^0\,Z^0$, and $b\overline{b}$, but not for $\tau^+ \tau^-$ and $\mu^+\mu^-$.  Therefore, we will not compute the supersaturation constraint for the leptonic final states.\footnote{The fact that $\tau^+ \tau^-$ and $\mu^+\mu^-$ final states contribute photons at energies near the dark matter mass also affects the shape constraints and accounts for the differences in the exclusion curves for $W^+W^-$, $Z^0\,Z^0$, and $b\overline{b}$ (see Figs.~\ref{fig:ShapeConstraintWW} and \ref{fig:ShapeConstraintBB}) and $\tau^+ \tau^-$ and $\mu^+\mu^-$ final states (see Figs.~\ref{fig:ShapeConstraintTT} and \ref{fig:ShapeConstraintMM}).}  Due to the similarity in shape between $W^+ W^-$ and $b \overline{b}$, we only present the supersaturation constraint for the former.

\begin{figure}[t] %  figure placement: here, top, bottom, or page
   \centering
   \includegraphics[width=.55 \textwidth]{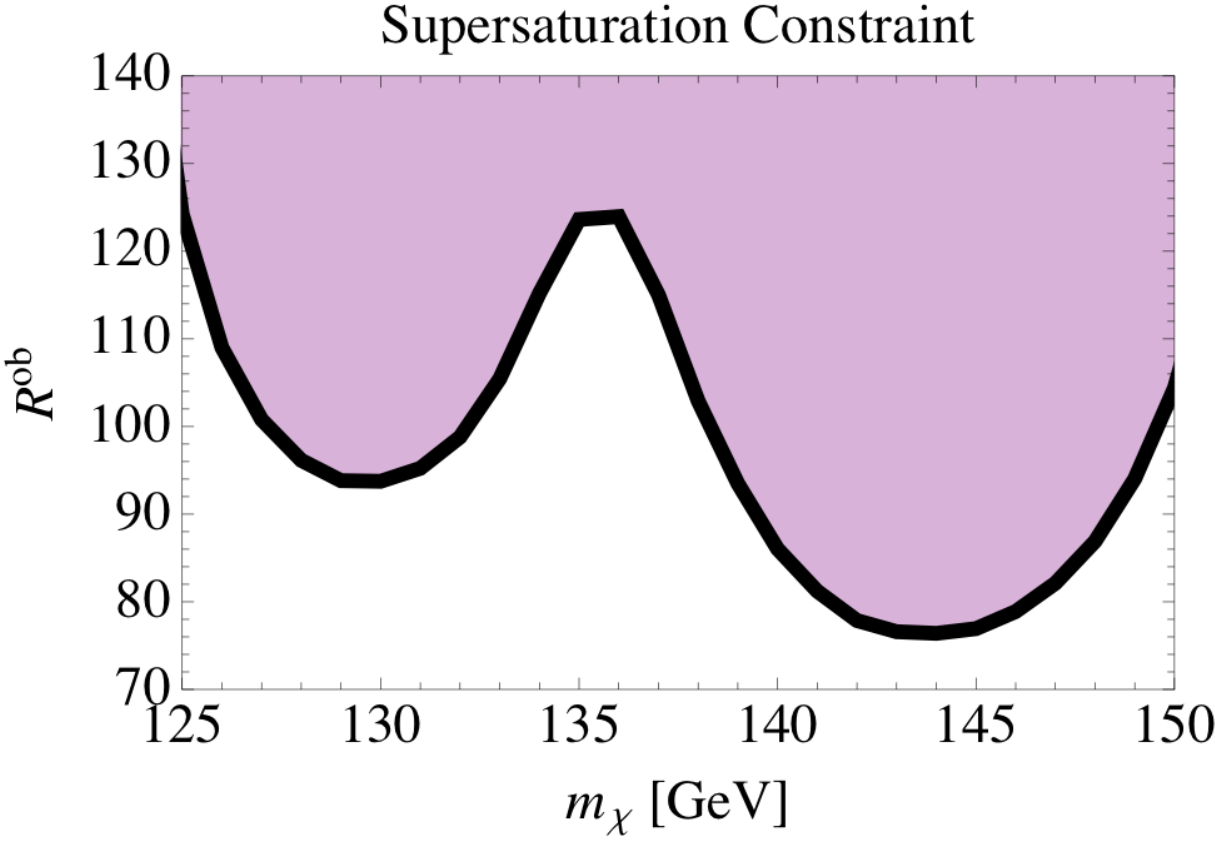}  
   \caption{The 95\% C.L. excluded region for $R^\mr{ob}$, as defined in \eref{eq:RObserved}, versus $m_\chi$ assuming annihilation into $W^+W^-$, $Z^0\, Z^0$ for the supersaturation analyses using the masked data set.  The plotted mass range corresponds to the 2 $\sigma$ best fit region.  For comparison, $R^{\text{th}}_{\text{wino}} \simeq 200$ and $R^{\text{th}}_{\text{Higgsino}} \simeq 700$.  Pure wino and Higgsino dark matter are clearly excluded, as discussed in Sec.~\ref{sec:MSSM}.}
   \label{fig:RContraintSupersaturation}
\end{figure}

In Fig.~\ref{fig:RContraintSupersaturation}, we show the 95\% C.L.~exclusion region for the supersaturation analysis.  The ``optimal bin" is $10-20$ GeV and contains 1104 photons.
To determine the exclusion region, we assume that the number of events in this bin are Poisson distributed and marginalize over the number of photons attributed to the $\gamma$-line signal (using a fit to the peak for data from 80--200 GeV).  For reference, there are about $29$ photons contributing to the $\gamma$-line at $m_\chi = 130 \mbox{ GeV}$ and $145 \mbox{ GeV}$ --- the results are qualitatively similar to those shown in Fig.~\ref{fig:FitGammaLine}.  To account for the change in effective area between the two bins, we apply a multiplicative correction of 1.06 to the number of peak photons; this is computed by taking the ratio of the effective area at 130 GeV and 15 GeV.  Using the spectrum obtained from \texttt{Pythia}, we determine the average number of photons per annihilation, $n_\mr{ann}^\gamma$, for the process $\chi \chi \rightarrow W^+W^-$ in the range 10--20 GeV.   For $m_\chi = 130$ GeV (145 GeV), $n_\mr{ann}^\gamma = 0.67$ (0.78).  

Requiring that dark matter annihilations to $W^+W^-$ do not supersaturate the data at the 95\% C.L.~constrains $R^\mr{ob} > 94$ (77) for $m_\chi = 130 \mbox{ GeV}$ (145 GeV).  Note that the constraint weakens at the edges of the considered mass range; at these masses, a line does not provide a good fit to the data and therefore the best fit value of $\NGG+\NGZ$ goes to zero, causing $R^\mr{ob}$ to become larger.  To good approximation, the limit on $R^\mr{ob}$ is the same for the $b\,\overline{b}$ final state.

Sec.~\ref{sec:MSSM} will show that there is a lower bound of $R^\mr{th} \gtrsim 200$ in the MSSM for annihilations to $W^+W^-$ and $Z^0 Z^0$.  Clearly, the supersaturation constraint robustly rules out this entire parameter space.

\subsection{Constraint Utilizing Shape Information}\label{sec:ShapeConstraint}

In this section, we present a complementary bound on $R^{\text{ob}}$ that utilizes the shape of the continuum spectrum.  The ratio $\Nann/(\NGG+\NGZ)$ is constrained by performing a log likelihood fit as described in Sec.~\ref{sec: fitsignal}.  For a given value of $\Nann/(\NGG+\NGZ)$ and $m_\chi$, we marginalize over $\alpha$, $\beta$, $\NGG$, and $\NGZ$.  This analysis is more constraining than the supersaturation results of Sec.~\ref{sec:SupersaturationConstraint}, but depends on the assumption that the $\gamma$ ray background is described by a single power law from 5--200 GeV.  

\begin{figure}[b] %  figure placement: here, top, bottom, or page
   \centering
   \begin{tabular}{cc}
\includegraphics[width=0.95\textwidth]{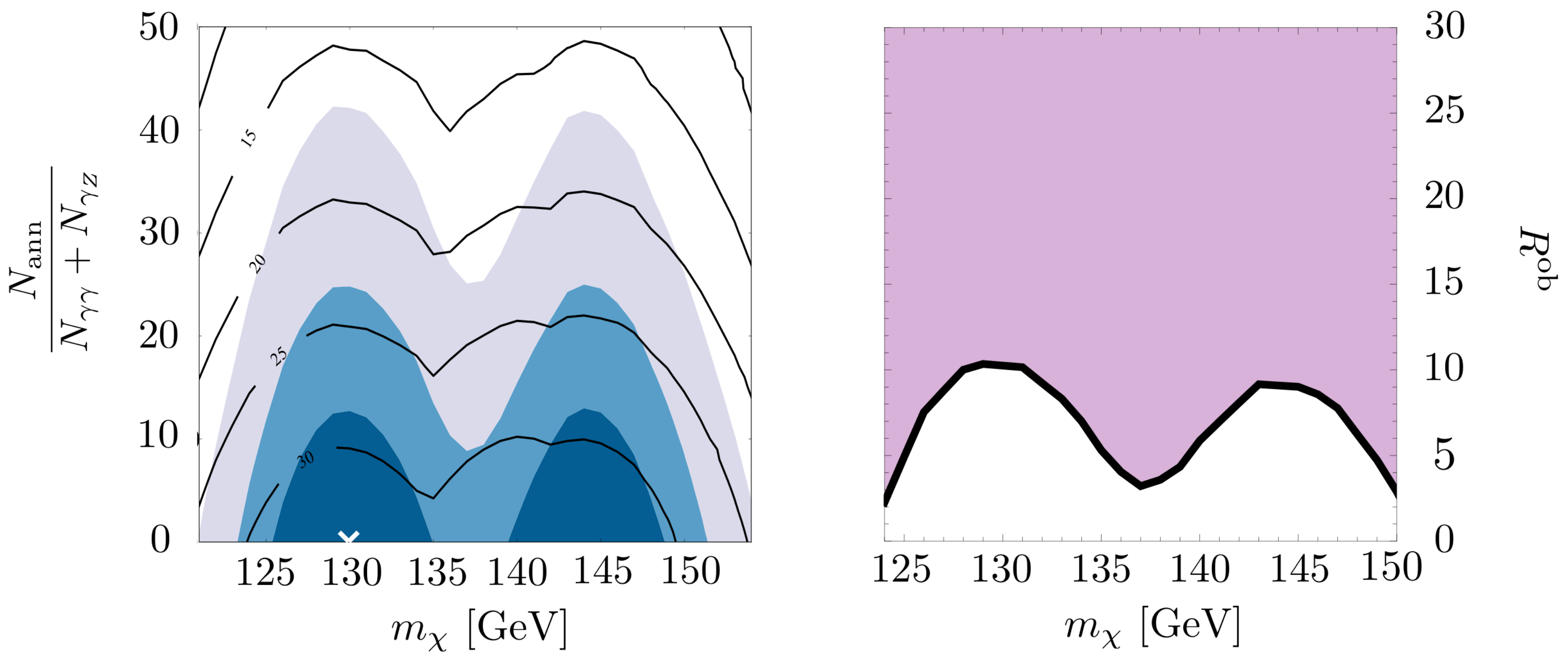}    \\
\includegraphics[width=.35\textwidth]{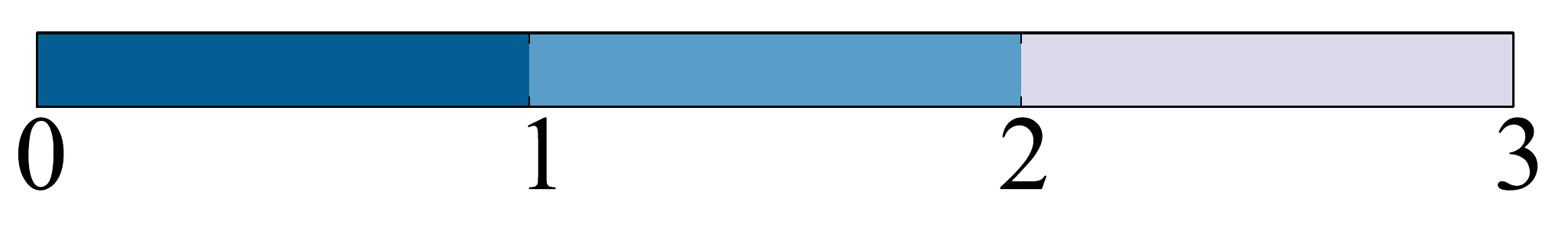} $\quad\quad\quad\quad\quad\quad\quad\quad\quad\quad\quad\quad\quad\quad\quad\quad\quad$
  \end{tabular} 
   \caption{(Left) The 1, 2, and 3 $\sigma$ confidence regions (filled contours) for $\Nann/(\NGG+\NGZ)$ as a function of mass for dark matter annihilation to $W^+W^-$.  The 1, 2, and 3 $\sigma$ contours refer to $\Delta \ln \mathcal{L} = 2.36, 4.86,$ and $8.13$ (4 d.o.f.).  The ratio $\NGZ/\NGG$ is allowed to freely vary for each point in the grid.  The solid black lines are the contours for $\NGG+\NGZ$.  The best fit point is marked with a cross at $m_{\chi} = 130$ GeV, $\thetaGZ = 0$, and $\Nann=0$.  The confidence regions for $Z^0 Z^0$ and $b\bar{b}$ are similar.   (Right)  The shape analysis constraint. The shaded region corresponds to parameters where the fit is 2 $\sigma$ or worse with respect to the best fit point.  This constraint is $\mathcal{O}$(10) stronger than the supersaturation constraint shown in Fig.~\ref{fig:RContraintSupersaturation}.  The constraints for $b\overline{b}$, $\tau^+\tau^-$, and $\mu^+\mu^-$ final states are provided in Appendix \ref{sec:AlternateFinalStates}.}
 \label{fig:ShapeConstraintWW}
\end{figure}

The best fit point is the same as  in \eref{eq:BestFitModel}, with $\Nann=0$.  The fact that the fit prefers no annihilation to $W^+W^-$ is not surprising.  Figure~\ref{fig:BestFit} shows that a single power law provides a remarkably good fit to the data between 5--100 GeV.  The filled contours in the left panel of Fig.~\ref{fig:ShapeConstraintWW} show the 1, 2, and 3 $\sigma$ confidence regions about the best fit point.  The black solid lines denote contours of $\NGG+\NGZ$.  There is some room for a non-zero annihilation contribution.  For these cases, the continuum spectrum explains the data below $\sim$15--20 GeV and the power law background becomes important at larger energies.  Typically, the best fit power law is shallower when $\Nann > 0$ than when $\Nann = 0$.

The 2 $\sigma$ confidence region for $\Nann/(\NGG+\NGZ)$ can be converted into a bound on $R^\mr{ob}$ by multiplying by $1/n^\gamma_\mr{ann}$ integrated over the appropriate energy range.  The result is given on the right in Fig.~\ref{fig:ShapeConstraintWW}, which shows the region excluded at 95\% C.L.~for $R^\mr{ob}$.  The maximum allowed value is $R^\mr{ob}_\mr{max}\simeq10$ for a mass of 129 GeV.  The entire range of $R^\mr{ob}$ is excluded outside the plotted range for $m_\chi$ because these masses do not provide a good fit to the data.

Electrons and positrons produced by dark matter annihilation can give additional contributions to the continuum from inverse Compton scattering (ICS) of the interstellar radiation field~\cite{Zavala:2011tt}. Neglecting this contribution is conservative for the supersaturation constraint, but one might wonder if the addition of ICS photons could improve the spectral correspondence between the model and the data for the shape constraint, hence weakening the limits.

The details of the ICS contribution depend on the diffusion and transport of charged particles in the inner Galaxy; however, for the $W^+ W^-$, $Z^0Z^0$, and $b\bar{b}$ annihilation channels, the ICS contribution is always subdominant.  To demonstrate this point, consider the case where \emph{all} the energy of the electrons and positrons is converted into upscattered photons via ICS (as opposed to other processes such as synchrotron), and furthermore where spatial diffusion can be neglected, with all scatterings occurring at the point of annihilation. These approximations likely overestimate the true ICS contribution; any significant energy density in magnetic fields will reduce the power in ICS photons in favor of synchrotron, and since we are examining the region where the line signal is concentrated, diffusion of electrons and positrons after annihilation should only dilute the corresponding ICS signal.

We obtain the electron and positron spectra produced by DM annihilation using \texttt{Pythia} as described previously, and then compute the spectrum of photons produced by repeated ICS (equivalently, the ICS photons produced by scattering on the steady-state electron+positron spectrum), under these simplifying approximations. Scattering of $\sim 100$ GeV photons on starlight approaches the Klein-Nishina regime, where the energy of the upscattered photons is comparable to the initial electron energy; consequently we use the full Klein-Nishina cross section and take into account that in this regime the electrons may lose a significant fraction of their energy in a single scattering (see \cite{1970RvMP...42..237B} for formulae and discussion). We employ the model for the interstellar radiation field at the Galactic Center used in \texttt{GALPROP} version 50p \cite{Porter:2005qx}; while there could be other radiation fields present, in order to dominate the ICS losses they would need to have an energy density exceeding that of the CMB or the starlight photons in the inner Galaxy, which seems implausible.

\begin{figure}[t]
\begin{center}
\includegraphics[width=0.6\textwidth]{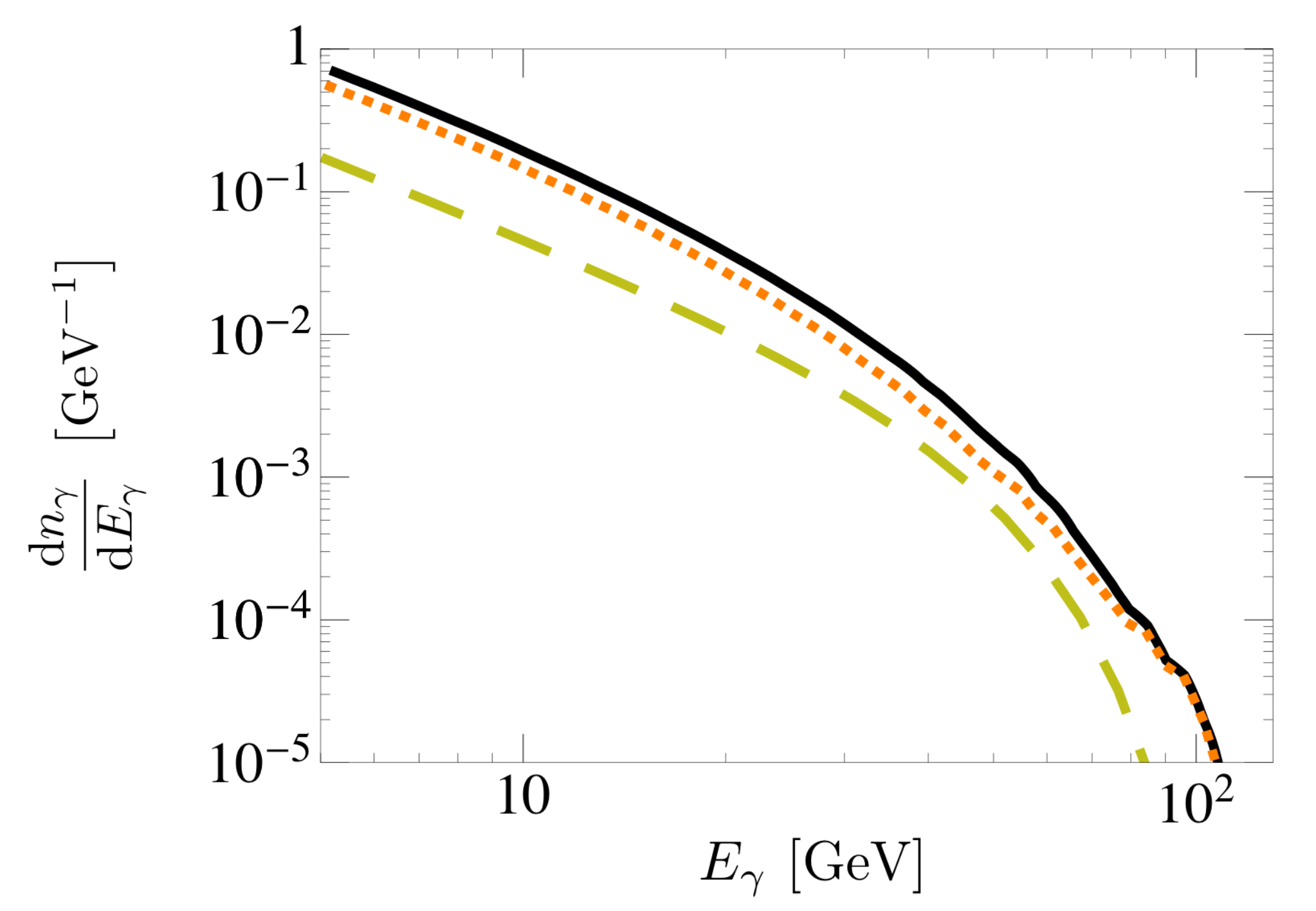}
\end{center}
\vskip -20pt
\caption{We show the spectrum of photons produced by the decays of $W^+ W^-$ from a single DM annihilation [orange, dotted], an estimate of the spectrum resulting from inverse Compton scattering of the associated electrons and positrons on the microwave, infrared and optical photons in the inner Galaxy [gold, dashed], and their sum [black, solid]. The dark matter mass is 130 GeV.  We see that for this final state, the photons from ICS are a negligible component of the full continuum photon spectrum in the energy range relevant to this study.}
\label{Fig:AnnWithIC}
\end{figure}

As shown in  Fig.~\ref{Fig:AnnWithIC}, for dark matter annihilations to $W^+ W^-$, the ICS spectrum is everywhere subdominant and adding it makes no significant difference to the continuum spectrum.  The same conclusion holds true for $Z^0Z^0$ and $b\bar{b}$. As a result, it is not necessary to include the ICS contribution in our fitting procedure. 

At low energies, the ICS contribution for leptonic final states computed using this simple estimate dominates over the directly produced photons.  We have explicitly checked the effect on our likelihood analysis.  The inclusion of the (likely overestimated) ICS contribution described above only tightens the limits; it makes the constraints more stringent while not shifting the best-fit points.  Thus, the bounds we present on leptonic final states are conservative. Because the best-fit value for $\Nann$ is non-zero for these final states, a more careful treatment of the ICS contribution may be worth future study.

%%%
\section{Implications for Neutralino Dark Matter}\label{sec:MSSM}
%%%

Now that we have explored the constraints on the dark matter scenario with both line and continuum contributions, this section will be devoted to the implications for MSSM neutralino dark matter.  The MSSM neutralino is a linear combination of bino $\widetilde{B}$, wino $\widetilde{W}$, up-type Higgsino $\widetilde{H}_u$, and down-type Higgsino $\widetilde{H}_d$.  Its properties are controlled by four parameters that determine the mass matrix: the bino mass $M_1$, the wino mass $M_2$, the Higgsino mass $\mu$, and $\tan \beta$.  The \texttt{DarkSUSY} program version 5.0.5 \cite{Gondolo:2004sc} is used to compute all relevant annihilation cross sections \cite{Bergstrom:1997fh, Ullio:1997ke, Bern:1997ng, Drees:1992am}.  The conventions for the neutralino mass matrix are the same as in \cite{Martin:1997ns}; we do \emph{not} impose any priors on the relic density.  

To understand the range of possible values for $\sigmaGG$, $\sigmaGZ$, and $\sigmaann$ in the MSSM, we begin by analyzing the pure gaugino eigenstates for $m_\chi = 130 \mbox{ GeV}$.  In the decoupling limit ($m_A \rightarrow \infty$) with heavy sfermions, the pure bino is inert and only pure wino and Higgsino states have non-negligible annihilation cross sections:
\bea
\mbox{\large\bf Wino:} &\quad\quad&
 \begin{array}{l}
\sigmaGG v \simeq 2.5 \times 10^{-27} \cms\\
\sigmaGZ v \simeq 1.4 \times 10^{-26} \cms\\
\sigmaann v \simeq \sigmaWW v \simeq 4.0 \times 10^{-24} \cms
\end{array} 
\,\,\,\quad\quad\quad\quad\Longrightarrow \quad R^\mr{th}  = 210;\label{eq:PureWino}\\
&&\nonumber\\
\mbox{and}\nonumber\\
&&\nonumber\\
\mbox{\large\bf Higgsino:} &\quad\quad&
\begin{array}{l}
\sigmaGG v \simeq 1.1 \times 10^{-28} \cms\\
\sigmaGZ v \simeq 3.7 \times 10^{-28} \cms\\
\sigmaann v \simeq  \sigmaWW v+\sigmaZZ v \simeq 4.2 \times 10^{-25} \cms
\end{array}
\quad\Longrightarrow \quad R^\mr{th}  = 710.\label{eq:PureHiggsino}
\eea
For these two cases, the cross sections to $\gamma \gamma$ and $\gamma Z^0$ are dominated by loops involving the charginos.  Annihilations to $W^+ W^-$ are due to $t$-channel chargino exchange.  The Higgsino also has a non-trivial cross section to $Z^0Z^0$ due to the presence of a light Higgsino-like second neutralino.  The pure Higgsino and pure wino are clearly ruled out by the constraints presented in Fig.~\ref{fig:RContraintSupersaturation}.

\begin{figure}[tb]
\begin{center}
\includegraphics[width=0.53\textwidth]{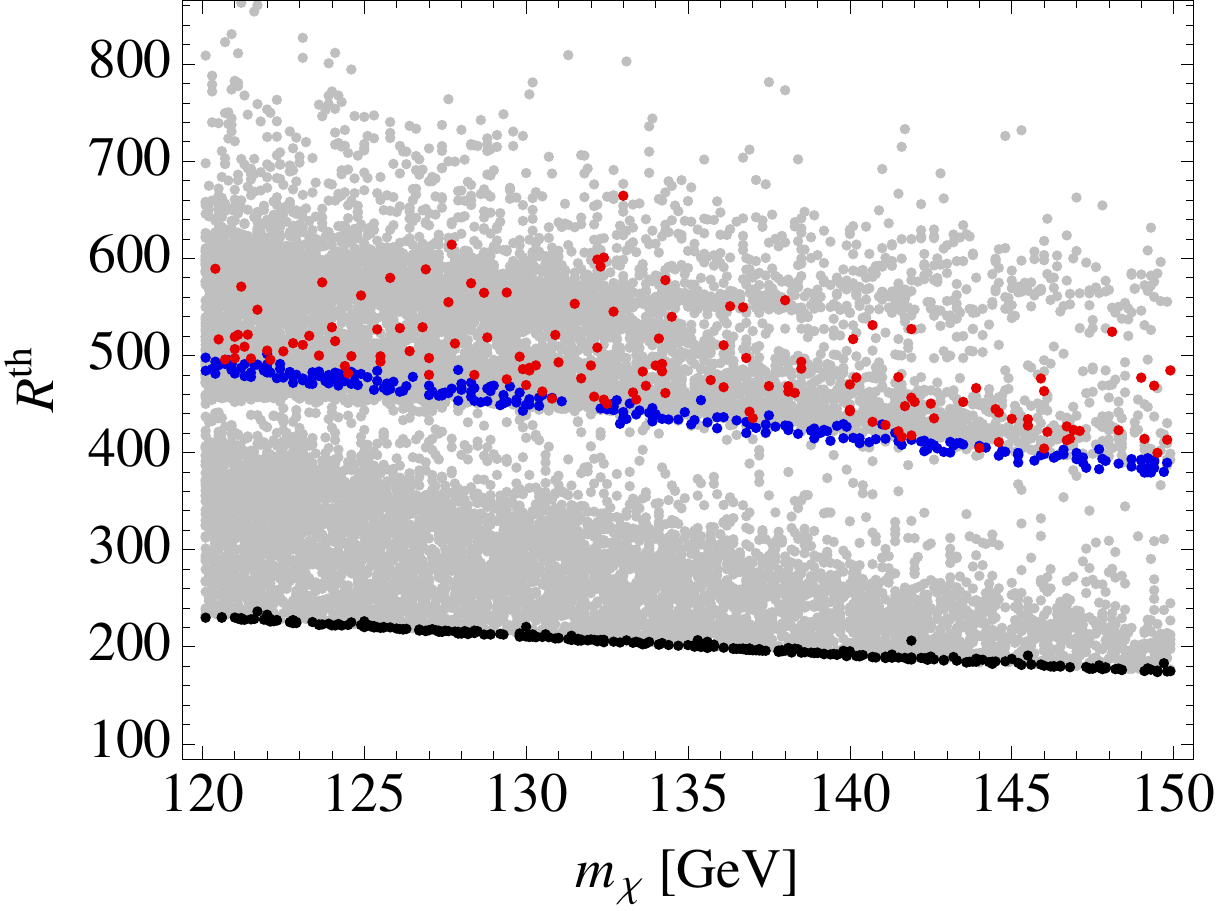}
\end{center}
\vskip -15pt
\caption{We have plotted $R^\mr{th}  = \sigma_\mr{ann}/(2\, \sigmaGG+\sigmaGZ)$ as a function of the neutralino mass.  Points with wino fraction $|Z_W|^2 > 0.99$ are plotted in black, points with Higgsino fraction $|Z_{H_u}|^2 + |Z_{H_d}|^2  > 0.99$ and $\tan \beta \geq 5$ ($\tan \beta < 5$) are plotted in blue (red), and points with $\mr{max}(\sigmaGG,\,1/2\,\sigmaGZ) > 10^{-32}\cms$ are plotted in grey.  We find a robust lower bound on $\sigma_\mr{ann}/(2\, \sigmaGG+\sigmaGZ)$ as a function of mass for neutralinos with a large enough annihilation rate to $\gamma\gamma$ and/or $\gamma Z^0$ to explain the 130 GeV line.  For all points in this plot $\sigma_\mr{ann}$ is dominated by some combination of $\sigmaWW$ and $\sigmaZZ$; the limits derived in Sec.~\ref{sec:ContConstraint} for $\chi \chi \rightarrow W^+ W^-$ are relevant.}
\label{Fig:WWtoGGRatioVsMass}
\end{figure}

To explore the case where the neutralino is a non-trivial admixture, we scan $M_1,M_2 > 100\mbox{ GeV}, |\mu| < 1 \mbox{ TeV}$, and randomize the sign of $\mu$, while keeping $m_A=m_{\tilde{f}}=3 \mbox{ TeV}$.  The range of possible $R^\mr{th}$ are shown in Fig.~\ref{Fig:WWtoGGRatioVsMass} for dark matter masses between 120  and 150$\mbox{ GeV}$.  Points with wino fraction $|Z_W|^2 > 0.99$ are plotted in black, points with Higgsino fraction $|Z_{H_u}|^2 + |Z_{H_d}|^2  > 0.99$ and $\tan \beta \geq 5$ ($\tan \beta < 5$) are plotted in blue (red), and points with $\mr{max}(\sigmaGG,\,1/2\,\sigmaGZ) > 10^{-32}\cms$ are plotted in gray.  This lower bound on the cross section eliminates points that do not have a large enough cross section to explain the line; it is an extremely conservative choice, being roughly five orders of magnitude below the cross section obtained by~\cite{Weniger:2012tx}.

Neutralinos that are dominantly wino or Higgsino give a sharp prediction for $R^\mr{th}$ as a function of mass (for moderate to large $\tan \beta$).  In particular, Higgsinos have a larger $R^\mr{th}$ than winos because $\sigma_\mr{ann}$ includes an $\mathcal{O}(1)$ contribution from $Z^0Z^0$ final states in addition to $W^+W^-$, which is the only final state for the wino case.  Note that the Higgsino-like prediction for $R^\mr{th} $ is different than the one given in \eref{eq:PureHiggsino} due to the finite scan range for $M_2$.  Because the wino cross section to $\gamma Z^0$ is larger than that for Higgsinos by two orders of magnitude, even a 1\% wino admixture can change the annihilation cross section for the Higgsino-like points non-trivially.

The large spread in $R^\mr{th}$ for Higgsinos with small $\tan \beta$ arises from an interesting effect in the neutralino mass matrix.  When $\tan \beta = 1$, one of the Higgsino states is always pure and when $\mu<0$, this state is the lightest Higgsino.  In this case, $M_2$ can be small while still maintaining the purity of the Higgsino.  Consequently, chargino and second neutralino mixing becomes non-trivial, lifting these masses above $\mu$.    The cross sections to $\gamma \gamma$ and $\gamma Z^0$ are suppressed faster than those for annihilation to $W^+ W^-$ and $Z^0Z^0$, which is why the red points are above the blue.  Fig.~\ref{Fig:WWtoGGRatioVsMass}  demonstrates that there is a lower bound on $R^\mr{th} $ in the decoupling limit with heavy sfermions, which is excluded by the constraint in  Fig.~\ref{fig:RContraintSupersaturation}.

If we relax the assumptions on the masses of the sfermions and additional Higgs bosons, $\sigmaGG$ and $\sigmaGZ$ change at the percent level and $\sigmaann$ is affected even less.  Even when parameters are tuned so that the neutralino annihilation is dominated by the $A$ resonance, the increase in the total annihilation cross section is orders of magnitude more then the increase in the rate for $\gamma\gamma$ or $\gamma Z^0$.  Hence, the lower bound on $R^\mr{th} $ for dominantly wino/Higgsino neutralinos is robust.

There is one final option to explore.  When the sfermions are light, the bino is no longer inert.  In particular, for $m_\chi = 130$ GeV and the slepton mass $m_{\tilde{\ell}} = 200\GeV$,\footnote{Note we are neglecting the possibility of a Sommerfeld enhanced annihilation to photons, which can happen when the sfermion mass is very close to the neutralino mass.  We thank Hai-Bo Yu for pointing out this possibility to us.  We are also omitting the effects of internal bremsstrahlung.} 
\bea
\mbox{{\large\bf Bino}:} &\quad&
\begin{array}{l}
\sigmaGG v \simeq \mbox{few} \times 10^{-30} \cms;\\
\sigmaGZ v \simeq \mbox{few} \times 10^{-31}  \cms;\\
\sigmaann v \simeq   \sigma_{\ell\bar{\ell}}\, v \simeq \mbox{few} \times 10^{-27}  \cms.
\end{array}\label{eq:PureBino}
\quad\Longrightarrow \quad R^\mr{th}  \sim 10^3.
\eea
The results of Figs.~\ref{fig:ShapeConstraintTT} and \ref{fig:ShapeConstraintMM}, which are relevant for lepton final states (see Appendix \ref{sec:AlternateFinalStates}), exclude this possibility.  However, the bino is not a good candidate for an even simpler reason.  Recall that the value for $\sigmaGG v$ found in \cite{Weniger:2012tx} for the Einasto profile is $1.3\times 10^{-27} \cms$.  Even allowing the large uncertainty in the shape of the profile at the Galactic Center, it is implausible that the cross sections given in \eref{eq:PureBino} are large enough to yield the observation.  As a result, the pure bino is not a good candidate to explain the $\gamma$-line.

These arguments along with Fig.~\ref{Fig:WWtoGGRatioVsMass} demonstrate that there is a lower bound on $\sigmaWW/(2\, \sigmaGG+\sigmaGZ)$ for neutralinos in the MSSM that have large enough $\sigmaGG$ and/or $\sigmaGZ$ to be consistent with the data.  Combined with the results of Sec.~\ref{sec:ContConstraint}, we see that the neutralino is excluded as an explanation of the 130 GeV \emph{Fermi} line.

\section{Conclusions}\label{sec:Discussion}

This paper presents constraints on the continuum photon spectrum for any dark matter candidate used to explain the 130 GeV line in the \emph{Fermi} data.  Many models have been proposed to explain this feature~\cite{Weiner:2012cb, Kang:2012bq, Das:2012ys, Chu:2012qy, Buckley:2012ws, Acharya:2012dz, Lee:2012bq, Kyae:2012vi, Choi:2012ap, Cline:2012nw, Dudas:2012pb, Feng:2012gs, Jackson:2009kg}, but all must now satisfy the strong requirements on continuum annihilation derived here.  In particular, we constrain the ratio of continuum photons versus monochromatic line photons, a quantity that is independent of astrophysical uncertainties.  The limit is strong enough to exclude typical models where the dark matter annihilates to $W^+W^-$ at tree level and into $\gamma Z^0$ and $\gamma \gamma$ at loop level.  In particular, this rules out the neutralino explanation of the 130 GeV $\gamma$ line.  

For $W^+W^-$, $Z^0Z^0$, and $b\bar{b}$ annihilation, the data prefers no continuum contribution.  For leptonic final states such as $\mu^+\mu^-$ and $\tau^+\tau^-$, a small continuum contribution for a 145 GeV dark matter candidate provides the best fit to the data.  Although the best fit models prefer little to no continuum, a model with line, background, and continuum contribution can still provide a good fit relative to a null power-law background.  For example, the ratio of continuum photons from $W^+W^-$ to line photons can be as large as $\sim 30$ and still provide 4 $\sigma$ improvement over the null model for 130 or 145 GeV dark matter.  A 3 $\sigma$ improvement over null is still achievable with a ratio about as large as $50$.

Figure \ref{fig:RContraintSupersaturation} shows that the more aggressive shape constraint on continuum photons places a limit on the ratio of $W^+W^-, Z^0Z^0$ to $\gamma\gamma$ annihilations of $\mathcal{O}(10)$.  Any model that is proposed to explain the $\gamma$ line must therefore have suppressed annihilation to these final states.  As one simple example of a model that could be consistent with these constraints, suppose the dark matter has only one interaction --- a Yukawa coupling to a new fermion $f$ and scalar $\widetilde{f}$ that carry electroweak quantum numbers.  If the mass of $f$ and $\widetilde{f}$ are larger than the dark matter mass, the leading annihilation is loop suppressed, and the ratios of $W^+W^-, Z^0Z^0$ to $\gamma\gamma$ annihilations are $\mathcal{O}(1)$.  While a more detailed computation could determine the allowed range of quantum numbers/couplings for $f$ and $\widetilde{f}$, this toy example illustrates a direction one might take when model building for the \emph{Fermi} line.

%Although a more detailed computation is needed to determine the allowed range of quantum numbers/couplings for $f$ and $\widetilde{f}$, this toy example illustrates some of the constraints that must be considered when model building for the \emph{Fermi} line.

There are reasons to be skeptical that this signal is the result of dark matter annihilations.
The 130 GeV $\gamma$ line was also observed along the Galactic plane along with evidence for other peaks in the gamma ray spectrum, implying that perhaps a more substantial look elsewhere correction should be applied~\cite{Boyarsky:2012ca}.  Furthermore, a slight excess at 130 GeV was observed in data taken from the Earth's albedo, which could point to the presence of an unidentified instrumental bias for reconstructing photons at this energy \cite{Su:2012ft}.  The question remains whether the feature observed in \cite{Weniger:2012tx} is a genuine dark matter signal, a systematic effect, or a new background~\cite{Profumo:2012tr}. 

If the 130 GeV line survives further scrutiny and is not a systematic or unknown background, then the statistical uncertainty on the line signal will be reduced as  \emph{Fermi} takes more data.  This will tighten the constraint on the annihilation cross sections.  In particular, the constraint on the ratio of $\gamma Z^0$ to $Z^0Z^0$ could approach $\mathcal{O}$(1).  From the effective field theory point of view, it is difficult to understand how annihilations to photons will not also be accompanied by annihilations to $Z^0$ bosons.  Hence, the updated versions of the constraints presented here could provide a significant challenge for model building a dark matter interpretation of the 130 GeV $\gamma$ line.
\vskip -10pt
\section*{Note Added}
\vskip -10pt
At the time of completion of this manuscript, \cite{Buchmuller:2012rc} appeared, which explores constraints on neutralino dark matter using the continuum spectrum from 20--200 GeV.  Our results are in agreement.
\vskip -15pt
\section*{Acknowledgements}
\vskip -10pt
We thank E.~Albin, N.~Arkani-Hamed, K.~Blum, M.~Peskin, N.~Weiner, C.~Weniger, and D.~Whiteson for useful discussions.  
TC is supported by the US Department of Energy under contract number DE-AC02-76SF00515.
ML is supported by the Simons Postdoctoral Fellows Program and the U.S. National Science Foundation, grant NSF-PHY-0705682, the LHC Theory Initiative.  TRS is supported by NSF grants PHY-0969448 and AST-0807444.  JGW is supported by the US Department of Energy under contract number DE-AC02-76SF00515.

\pagebreak

\appendix

\section{\emph{Fermi} Data Counts per Bin}
\label{sec:CountsPerBin}
\label{AppA}

\setlength{\tabcolsep}{6pt}
\begin{table}[h!]
\begin{tabular}{ccc|ccc|ccc|ccc}
$E_\text{min}$ & $E_{\text{max}}$ & Counts & $E_\text{min}$ & $E_{\text{max}}$ & Counts & $E_\text{min}$ & $E_{\text{max}}$ & Counts & $E_{\text{min}}$&$E_{\text{max}}$ & Counts \\
\hline
 5.05 & 5.20 & 221 & 12.64 & 13.00 & 51 & 31.62 & 32.54 & 12 & 79.12 &
   81.42 & 3 \\
 5.20 & 5.35 & 223 & 13.00 & 13.38 & 52 & 32.54 & 33.49 & 14 & 81.42 &
   83.79 & 3 \\
 5.35 & 5.50 & 216 & 13.38 & 13.77 & 46 & 33.49 & 34.46 & 5 & 83.79 & 86.23
   & 4 \\
 5.50 & 5.66 & 214 & 13.77 & 14.17 & 37 & 34.46 & 35.46 & 12 & 86.23 &
   88.73 & 1 \\
 5.66 & 5.83 & 204 & 14.17 & 14.58 & 42 & 35.46 & 36.49 & 13 & 88.73 &
   91.31 & 3 \\
 5.83 & 6.00 & 179 & 14.58 & 15.01 & 42 & 36.49 & 37.55 & 16 & 91.31 &
   93.97 & 1 \\
 6.00 & 6.17 & 185 & 15.01 & 15.44 & 45 & 37.55 & 38.65 & 9 & 93.97 & 96.70
   & 2 \\
 6.17 & 6.35 & 180 & 15.44 & 15.89 & 25 & 38.65 & 39.77 & 9 & 96.70 & 99.51
   & 1 \\
 6.35 & 6.54 & 168 & 15.89 & 16.36 & 37 & 39.77 & 40.93 & 12 & 99.51 &
   102.41 & 1 \\
 6.54 & 6.73 & 166 & 16.36 & 16.83 & 32 & 40.93 & 42.12 & 18 & 102.41 &
   105.38 & 2 \\
 6.73 & 6.92 & 150 & 16.83 & 17.32 & 35 & 42.12 & 43.34 & 6 & 105.38 &
   108.45 & 4 \\
 6.92 & 7.12 & 141 & 17.32 & 17.82 & 31 & 43.34 & 44.60 & 8 & 108.45 &
   111.60 & 1 \\
 7.12 & 7.33 & 148 & 17.82 & 18.34 & 34 & 44.60 & 45.90 & 9 & 111.60 &
   114.85 & 5 \\
 7.33 & 7.54 & 131 & 18.34 & 18.88 & 21 & 45.90 & 47.23 & 8 & 114.85 &
   118.19 & 0 \\
 7.54 & 7.76 & 128 & 18.88 & 19.42 & 32 & 47.23 & 48.61 & 9 & 118.19 &
   121.62 & 1 \\
 7.76 & 7.99 & 111 & 19.42 & 19.99 & 21 & 48.61 & 50.02 & 7 & 121.62 &
   125.16 & 3 \\
 7.99 & 8.22 & 123 & 19.99 & 20.57 & 28 & 50.02 & 51.47 & 8 & 125.16 &
   128.80 & 8 \\
 8.22 & 8.46 & 109 & 20.57 & 21.17 & 24 & 51.47 & 52.97 & 5 & 128.80 &
   132.54 & 7 \\
 8.46 & 8.71 & 96 & 21.17 & 21.78 & 22 & 52.97 & 54.51 & 6 & 132.54 &
   136.40 & 4 \\
 8.71 & 8.96 & 92 & 21.78 & 22.42 & 23 & 54.51 & 56.09 & 4 & 136.40 &
   140.36 & 1 \\
 8.96 & 9.22 & 97 & 22.42 & 23.07 & 14 & 56.09 & 57.73 & 4 & 140.36 &
   144.44 & 4 \\
 9.22 & 9.49 & 90 & 23.07 & 23.74 & 15 & 57.73 & 59.40 & 10 & 144.44 &
   148.64 & 1 \\
 9.49 & 9.76 & 73 & 23.74 & 24.43 & 21 & 59.40 & 61.13 & 6 & 148.64 &
   152.97 & 1 \\
 9.76 & 10.05 & 76 & 24.43 & 25.14 & 21 & 61.13 & 62.91 & 3 & 152.97 &
   157.41 & 0 \\
 10.05 & 10.34 & 78 & 25.14 & 25.87 & 11 & 62.91 & 64.74 & 2 & 157.41 &
   161.99 & 0 \\
 10.34 & 10.64 & 64 & 25.87 & 26.62 & 18 & 64.74 & 66.62 & 5 & 161.99 &
   166.70 & 0 \\
 10.64 & 10.95 & 71 & 26.62 & 27.40 & 20 & 66.62 & 68.56 & 4 & 166.70 &
   171.55 & 1 \\
 10.95 & 11.27 & 75 & 27.40 & 28.20 & 14 & 68.56 & 70.55 & 3 & 171.55 &
   176.54 & 1 \\
 11.27 & 11.60 & 67 & 28.20 & 29.01 & 12 & 70.55 & 72.60 & 6 & 176.54 &
   181.67 & 1 \\
 11.60 & 11.93 & 62 & 29.01 & 29.86 & 11 & 72.60 & 74.71 & 4 & 181.67 &
   186.95 & 1 \\
 11.93 & 12.28 & 53 & 29.86 & 30.73 & 13 & 74.71 & 76.89 & 0 & 186.95 &
   192.39 & 2 \\
 12.28 & 12.64 & 51 & 30.73 & 31.62 & 14 & 76.89 & 79.12 & 7 & 192.39 &
   197.98 & 1 \\
   \end{tabular}
 \caption{ Counts of \texttt{Pass 7\_Version 6} \texttt{ULTRACLEAN} events within $3^{\circ}$ of the Galactic Center, binned in energy [GeV], with the inner $1^{\circ}$ masked.  Analysis cuts are as described in the text.}
\end{table}

\pagebreak

\setlength{\tabcolsep}{6pt}
\begin{table}[h!]
\begin{tabular}{ccc|ccc|ccc|ccc}
$E_\text{min}$ & $E_{\text{max}}$ & Counts & $E_\text{min}$ & $E_{\text{max}}$ & Counts & $E_\text{min}$ & $E_{\text{max}}$ & Counts & $E_{\text{min}}$&$E_{\text{max}}$ & Counts \\
\hline
5.05 & 5.20 & 315 & 12.64 & 13.00 & 70 & 31.62 & 32.54 & 14 & 79.12 &
   81.42 & 3 \\
 5.20 & 5.35 & 328 & 13.00 & 13.38 & 72 & 32.54 & 33.49 & 19 & 81.42 &
   83.79 & 3 \\
 5.35 & 5.50 & 312 & 13.38 & 13.77 & 69 & 33.49 & 34.46 & 8 & 83.79 & 86.23
   & 4 \\
 5.50 & 5.66 & 311 & 13.77 & 14.17 & 61 & 34.46 & 35.46 & 13 & 86.23 &
   88.73 & 1 \\
 5.66 & 5.83 & 297 & 14.17 & 14.58 & 58 & 35.46 & 36.49 & 18 & 88.73 &
   91.31 & 5 \\
 5.83 & 6.00 & 278 & 14.58 & 15.01 & 58 & 36.49 & 37.55 & 18 & 91.31 &
   93.97 & 1 \\
 6.00 & 6.17 & 265 & 15.01 & 15.44 & 52 & 37.55 & 38.65 & 11 & 93.97 &
   96.70 & 3 \\
 6.17 & 6.35 & 253 & 15.44 & 15.89 & 39 & 38.65 & 39.77 & 12 & 96.70 &
   99.51 & 1 \\
 6.35 & 6.54 & 244 & 15.89 & 16.36 & 47 & 39.77 & 40.93 & 12 & 99.51 &
   102.41 & 2 \\
 6.54 & 6.73 & 242 & 16.36 & 16.83 & 54 & 40.93 & 42.12 & 21 & 102.41 &
   105.38 & 4 \\
 6.73 & 6.92 & 212 & 16.83 & 17.32 & 48 & 42.12 & 43.34 & 7 & 105.38 &
   108.45 & 4 \\
 6.92 & 7.12 & 196 & 17.32 & 17.82 & 38 & 43.34 & 44.60 & 9 & 108.45 &
   111.60 & 3 \\
 7.12 & 7.33 & 201 & 17.82 & 18.34 & 42 & 44.60 & 45.90 & 12 & 111.60 &
   114.85 & 5 \\
 7.33 & 7.54 & 182 & 18.34 & 18.88 & 28 & 45.90 & 47.23 & 12 & 114.85 &
   118.19 & 0 \\
 7.54 & 7.76 & 193 & 18.88 & 19.42 & 41 & 47.23 & 48.61 & 13 & 118.19 &
   121.62 & 1 \\
 7.76 & 7.99 & 156 & 19.42 & 19.99 & 29 & 48.61 & 50.02 & 9 & 121.62 &
   125.16 & 3 \\
 7.99 & 8.22 & 179 & 19.99 & 20.57 & 38 & 50.02 & 51.47 & 10 & 125.16 &
   128.80 & 8 \\
 8.22 & 8.46 & 144 & 20.57 & 21.17 & 30 & 51.47 & 52.97 & 5 & 128.80 &
   132.54 & 9 \\
 8.46 & 8.71 & 135 & 21.17 & 21.78 & 28 & 52.97 & 54.51 & 10 & 132.54 &
   136.40 & 4 \\
 8.71 & 8.96 & 136 & 21.78 & 22.42 & 28 & 54.51 & 56.09 & 6 & 136.40 &
   140.36 & 1 \\
 8.96 & 9.22 & 133 & 22.42 & 23.07 & 21 & 56.09 & 57.73 & 5 & 140.36 &
   144.44 & 4 \\
 9.22 & 9.49 & 127 & 23.07 & 23.74 & 22 & 57.73 & 59.40 & 11 & 144.44 &
   148.64 & 2 \\
 9.49 & 9.76 & 114 & 23.74 & 24.43 & 31 & 59.40 & 61.13 & 6 & 148.64 &
   152.97 & 1 \\
 9.76 & 10.05 & 95 & 24.43 & 25.14 & 25 & 61.13 & 62.91 & 5 & 152.97 &
   157.41 & 0 \\
 10.05 & 10.34 & 111 & 25.14 & 25.87 & 15 & 62.91 & 64.74 & 2 & 157.41 &
   161.99 & 0 \\
 10.34 & 10.64 & 98 & 25.87 & 26.62 & 24 & 64.74 & 66.62 & 5 & 161.99 &
   166.70 & 0 \\
 10.64 & 10.95 & 94 & 26.62 & 27.40 & 27 & 66.62 & 68.56 & 5 & 166.70 &
   171.55 & 1 \\
 10.95 & 11.27 & 113 & 27.40 & 28.20 & 21 & 68.56 & 70.55 & 3 & 171.55 &
   176.54 & 1 \\
 11.27 & 11.60 & 89 & 28.20 & 29.01 & 19 & 70.55 & 72.60 & 7 & 176.54 &
   181.67 & 1 \\
 11.60 & 11.93 & 91 & 29.01 & 29.86 & 18 & 72.60 & 74.71 & 6 & 181.67 &
   186.95 & 3 \\
 11.93 & 12.28 & 72 & 29.86 & 30.73 & 16 & 74.71 & 76.89 & 2 & 186.95 &
   192.39 & 2 \\
 12.28 & 12.64 & 76 & 30.73 & 31.62 & 20 & 76.89 & 79.12 & 8 & 192.39 &
   197.98 & 1 \\
   \end{tabular}
 \caption{ Counts of \texttt{Pass 7\_Version 6} \texttt{ULTRACLEAN} events within $3^{\circ}$ of the Galactic Center, binned in energy [GeV].   Analysis cuts are as described in the text.}
\end{table}

\pagebreak

\section{Modeling the \emph{Fermi} Instrument Response Function}\label{sec:FermiIRF}
\label{AppB}

\emph{Fermi}'s Instrument Response Function describes the energy dispersion for a photon that hits the detector at a given energy and angle of incidence and can be modeled using the publicly available information in the Science Tools documentation.\footnote{http://fermi.gsfc.nasa.gov/ssc/data/analysis/}  The energy dispersion of the \emph{Fermi} LAT is defined as 
\begin{equation}
\frac{\delta E}{E} = \frac{E' - E}{E},
\end{equation}
where $E$ is the true energy of the event and $E'$ is the reconstructed energy.  For the fits, a scaled dispersion $x$ is used, where
\begin{equation}
x \equiv \frac{\delta E}{E \cdot S_D(E, \theta)}.
\label{eq: scaling}
\end{equation}
The scaling factor $S_D$ depends on the true energy $E$ and true incidence angle $\theta$ of the event.  It is fit with the function
\begin{equation}
S_D(E, \theta) = c_0 \Bigg(\log \frac{E}{\text{[MeV]}}\Bigg)^2 + c_1 (\cos\theta)^2 + c_2 \log \frac{E}{\text{[MeV]}}+ c_3 \cos\theta + c_4 \log  \frac{E}{\text{[MeV]}} \cos\theta +c_5.
\end{equation}
For front-converted events, $(c_0, c_1, c_2, c_3, c_4, c_5) = (0.0210, 0.0580, -0.207, -0.213, 0.042, 0.564)$. 
For back-converted events, $(c_0, c_1, c_2, c_3, c_4, c_5) = (0.0215, 0.0507, -0.220, -0.243, 0.065, 0.584)$. 

The energy dispersion depends on both the energy and incidence angle of the incoming photon.  Its functional form is 
\begin{equation}
D^{ij}(x) = 
   \left\{\begin{array}{l}
      N_L\, R\left(x, x_0^{ij}, \sigma^{ij}_L, \gamma_L\right)  \quad\quad \text{if } (x-x_0) < -\tilde{x}\\ 
       \\[-8pt]
      N_l\, R\left(x, x_0^{ij}, \sigma_l^{ij}, \gamma_l\right)  \,\,\,\,\quad\quad  \text{if } -\tilde{x} \leq (x-x_0) \leq 0  \\
       \\[-8pt]
      N_r\, R\left(x, x_0^{ij}, \sigma_r^{ij}, \gamma_r\right)  \,\,\quad\quad \text{if } 0 \leq (x-x_0) \leq \tilde{x} \\
       \\[-8pt]
      N_R\, R\left(x, x_0^{ij}, \sigma_R^{ij}, \gamma_R\right)  \quad\quad \text{if } (x-x_0) > \tilde{x} 
    \end{array}\right.
\end{equation}

where
\begin{equation}
R(x, x_0, \sigma, \gamma) = \frac{\gamma}{2^{1/\gamma}\Gamma(1/\gamma)\, \sigma} \exp \Bigg[ -\frac{1}{2} \Big| \frac{x-x_0}{\sigma} \Big|^{\gamma} \Bigg].
\end{equation}
The superscript $i$ denotes the energy bin and $j$ denotes the $\cos\theta$ bin.  The values for $\tilde{x}$ and the gammas are given by
\begin{equation}
(\tilde{x}, \gamma_L, \gamma_l, \gamma_r, \gamma_R) = (1.5, 0.6, 1.6, 1.6, 0.6).
\end{equation}
The normalization factors $N_L, N_l, N_r, N_R$ are chosen such that $D(x)$ is continuous and $\int_{-\infty}^{\infty} D(x) \mr{d}x=1$.  
The remaining five free parameters $(x_0, \sigma_L, \sigma_l, \sigma_r, \sigma_R)$ are given by fits to the data for 18 separate energy bins and 8 separate $\cos\theta$ bins, for the front and back-converted events separately.  

The goal is to obtain the distribution for $\delta E$, properly averaged over $\cos\theta$ as well as the front- and back-converted events.  For a given energy bin,
\begin{equation}
D^i_{\text{avg}} = \frac{ \sum_{j=1}^{N_{c_\theta}} \Big[ D^{ij}_{\text{front}}(x) X^j A_{\text{eff, front}}^{ij} + D^{ij}_{\text{back}}(x) X^j A_{\text{eff,back}}^{ij} \Big] }{\sum_{j=1}^{N_{c_\theta}} \Big[X^j A_{\text{eff,front}}^{ij}  + X^j A_{\text{eff,back}}^{ij}\Big]}, 
\end{equation}
where $N_{c_\theta}$ is the number of bins in $\cos\theta$, $X^j$ is the exposure (summed over pixels within $3^{\circ}$ of the Galactic Center), and $A_{\text{eff}}^{ij}$ is the effective area for the front and back converters of the tracker.  

Recall that $x$ depends on $\cos\theta$ through $S_D$.  Therefore, we substitute Eq.~(\ref{eq: scaling}) into the expression for $D^i(x)$, and evaluate $S_D$ at the median value of $\cos\theta$ in the appropriate bin.  If one specifies the true energy of the photon, $E$, where $E$ is within the $i^{\text{th}}$ energy bin, then 
\begin{equation}
x = \frac{\delta E}{E \cdot S_D\left( E, \cos\theta_{\text{med}}\right)}.
\end{equation}

\begin{figure}[tb]
\begin{center}
\includegraphics[width=0.5\textwidth]{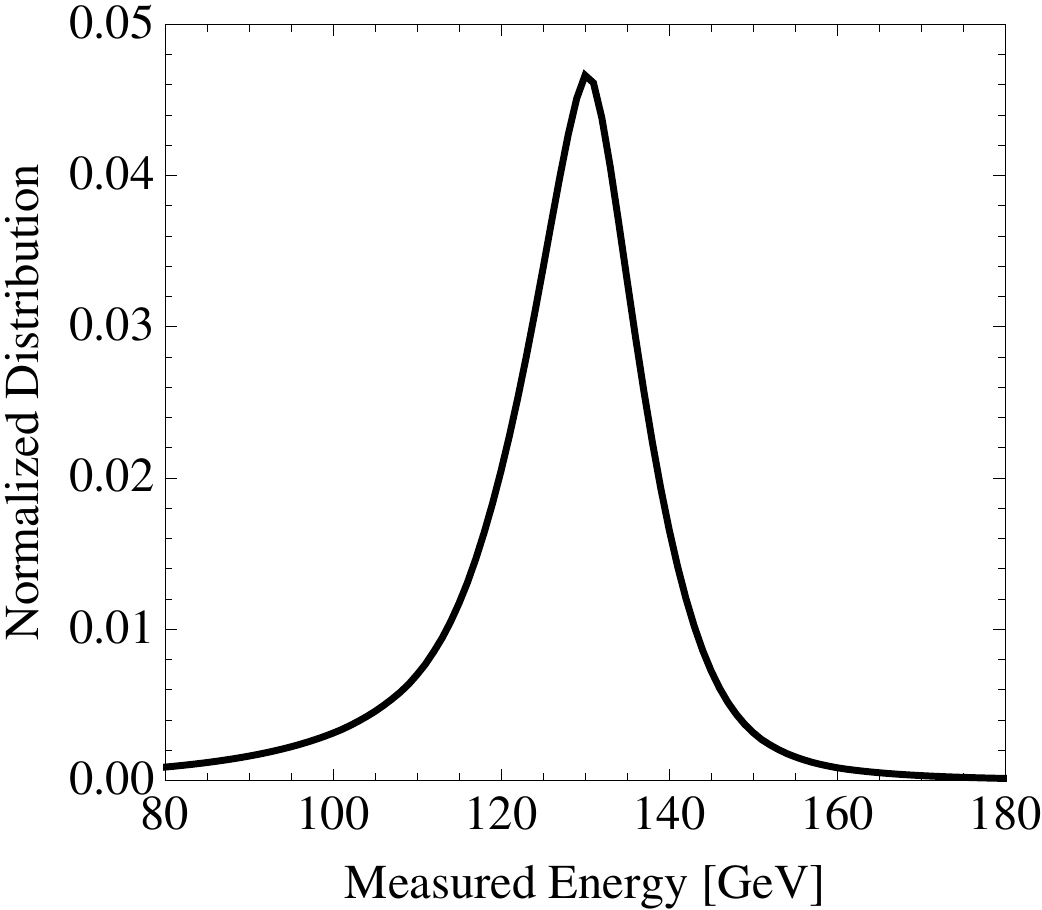}
\end{center}
\caption{The averaged energy dispersion $D^i(x)$ as a function of energy appropriate for a 130 GeV photon.}
\label{fig:EnergyDispersion}
\end{figure}

For reference, we plot the $D^i(x)$ for a 130 GeV photon in  Fig.~\ref{fig:EnergyDispersion}.

\pagebreak

\section{Constraints for Alternate Final States}\label{sec:AlternateFinalStates}
\label{AppC}
\begin{figure}[h!!] %  figure placement: here, top, bottom, or page
   \centering
 {\Large $\chi \chi \rightarrow b \overline{b}$}\\
\vskip 10 pt
  \includegraphics[width=0.9\textwidth]{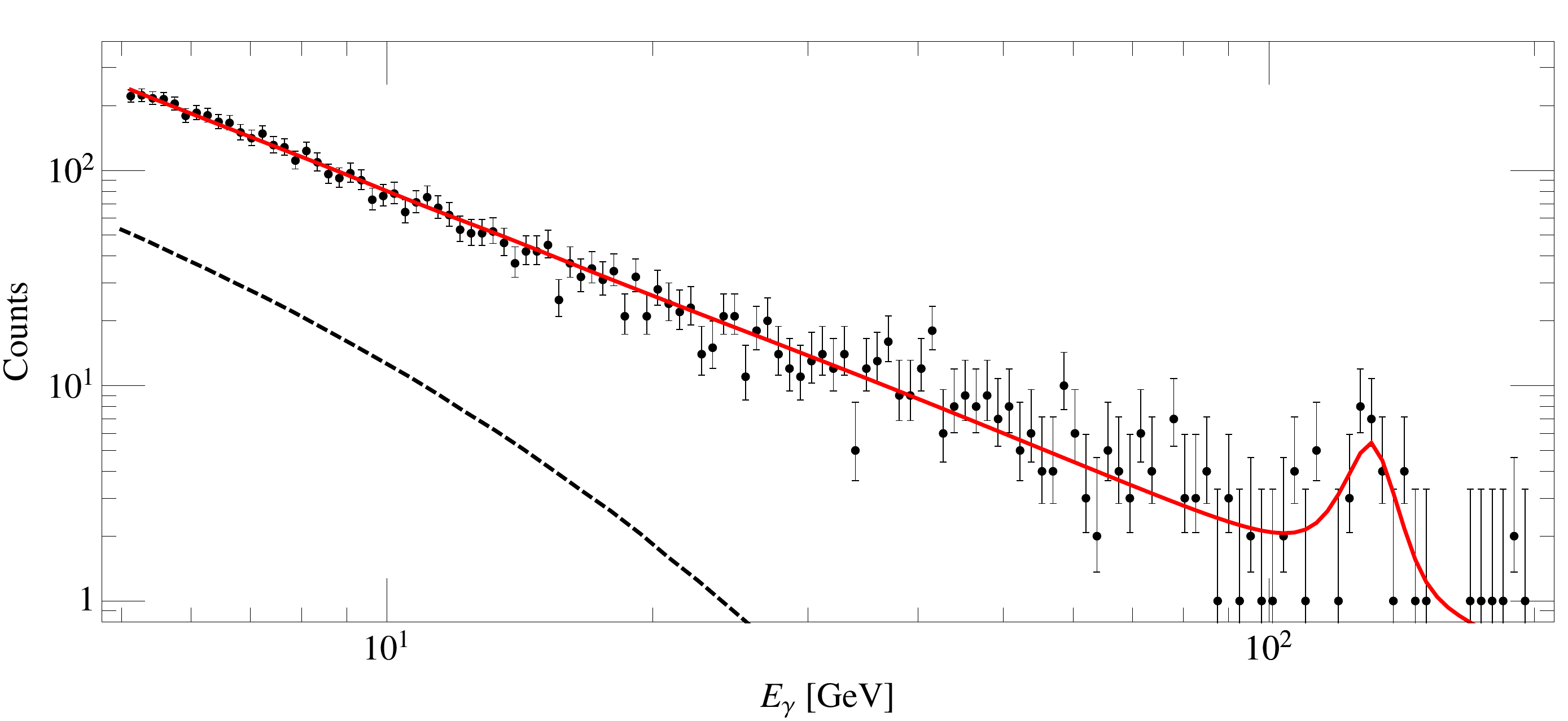}  \\
  \vskip 35 pt   
\includegraphics[width=1.05\textwidth]{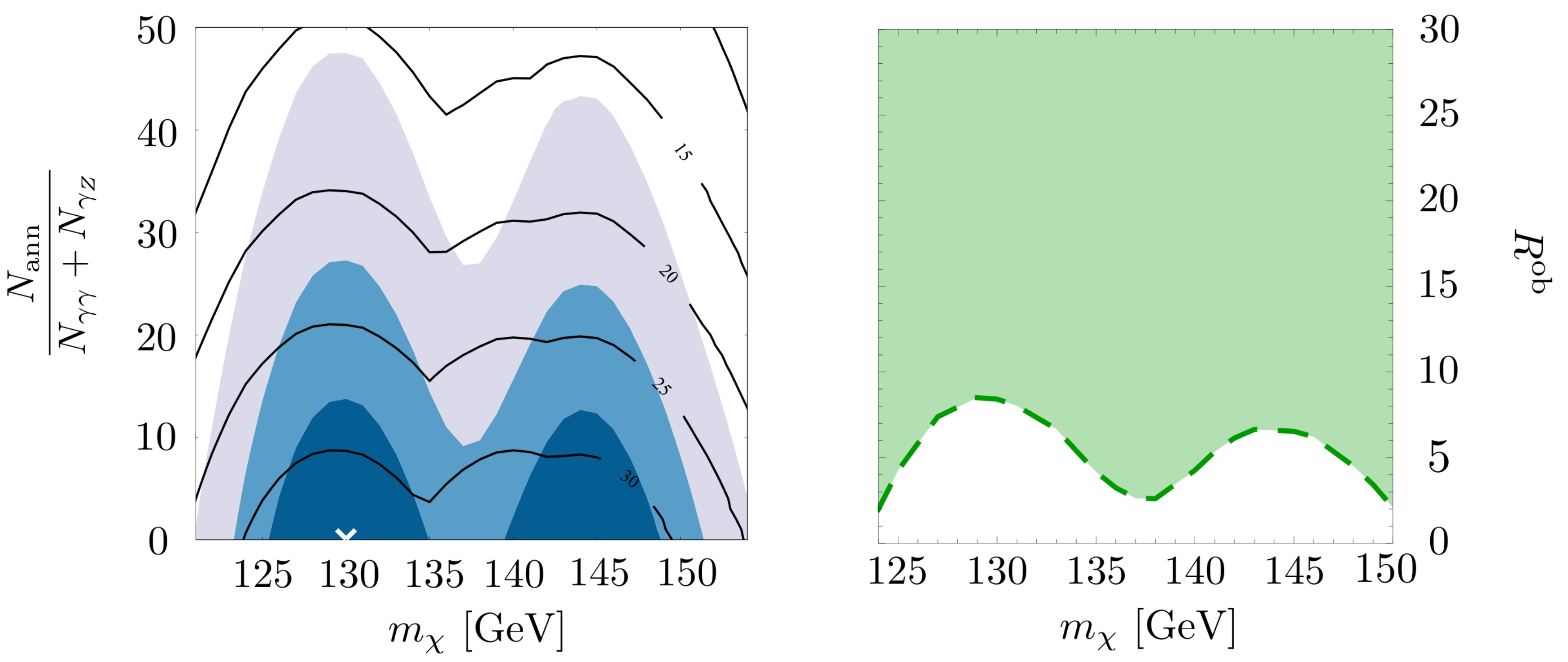}   
\includegraphics[width=.4\textwidth]{TemporaryDiagrams/colorbar.pdf} $\quad\quad\quad\quad\quad\quad\quad\quad\quad\quad\quad\quad\quad\quad\quad\quad\quad$ 
    \caption{The \emph{top} plot gives the photon counts within 3$^{\circ}$ degrees of the Galactic Center with the inner degree masked.  The solid red line shows the best fit model, which is given by the white cross in the bottom left plot.  This best fit point has  $\Nann = 0$; for reference the dashed black line shows the continuum spectrum for 130 GeV dark matter annihilating into $b\,\overline{b}$ with an arbitrary normalization.
On the \emph{bottom left}, we show 1, 2, and 3 $\sigma$ confidence regions (filled contours) for $\Nann/(\NGG+\NGZ)$ as a function of mass for dark matter annihilation to $b\,\overline{b}$.  The ratio $\NGZ/\NGG$ is allowed to freely vary for each point in the grid.  The solid black lines are the contours for $\NGG+\NGZ$.  The best fit point is marked with a cross at $m_{\chi} = 130$ GeV, $\thetaGZ = 0$, and $\Nann=0$.  On the \emph{bottom right}, we show the shape analysis constraint. The shaded region corresponds to parameters where the fit is 2 $\sigma$ or worse with respect to the best fit point.}
\label{fig:ShapeConstraintBB}
\end{figure}

\newpage

\begin{figure}[h!!] %  figure placement: here, top, bottom, or page
   \centering
 {\Large $\chi \chi \rightarrow \tau^+ \tau^-$}\\
\vskip 10 pt
  \includegraphics[width=.9 \textwidth]{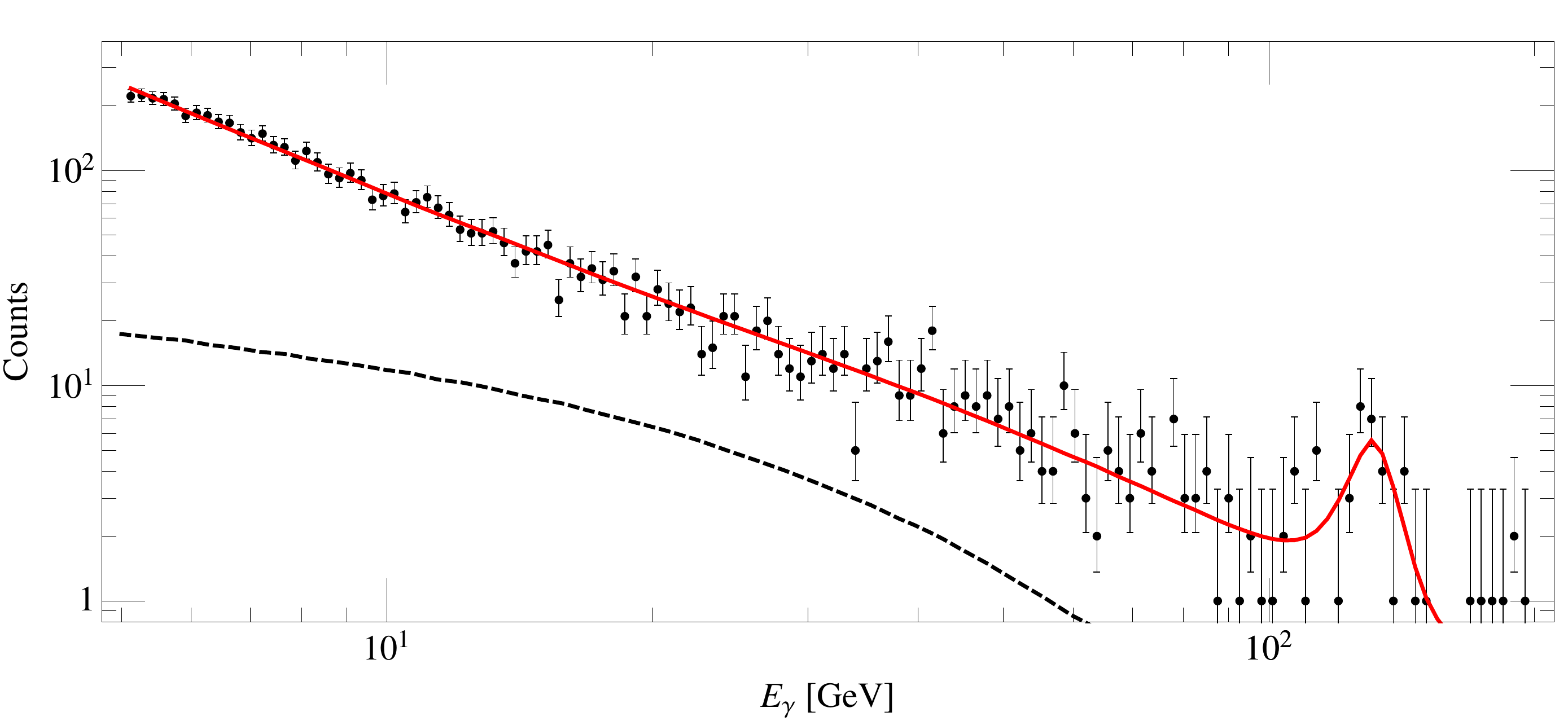}  \\
  \vskip 35 pt
\includegraphics[width=1.05\textwidth]{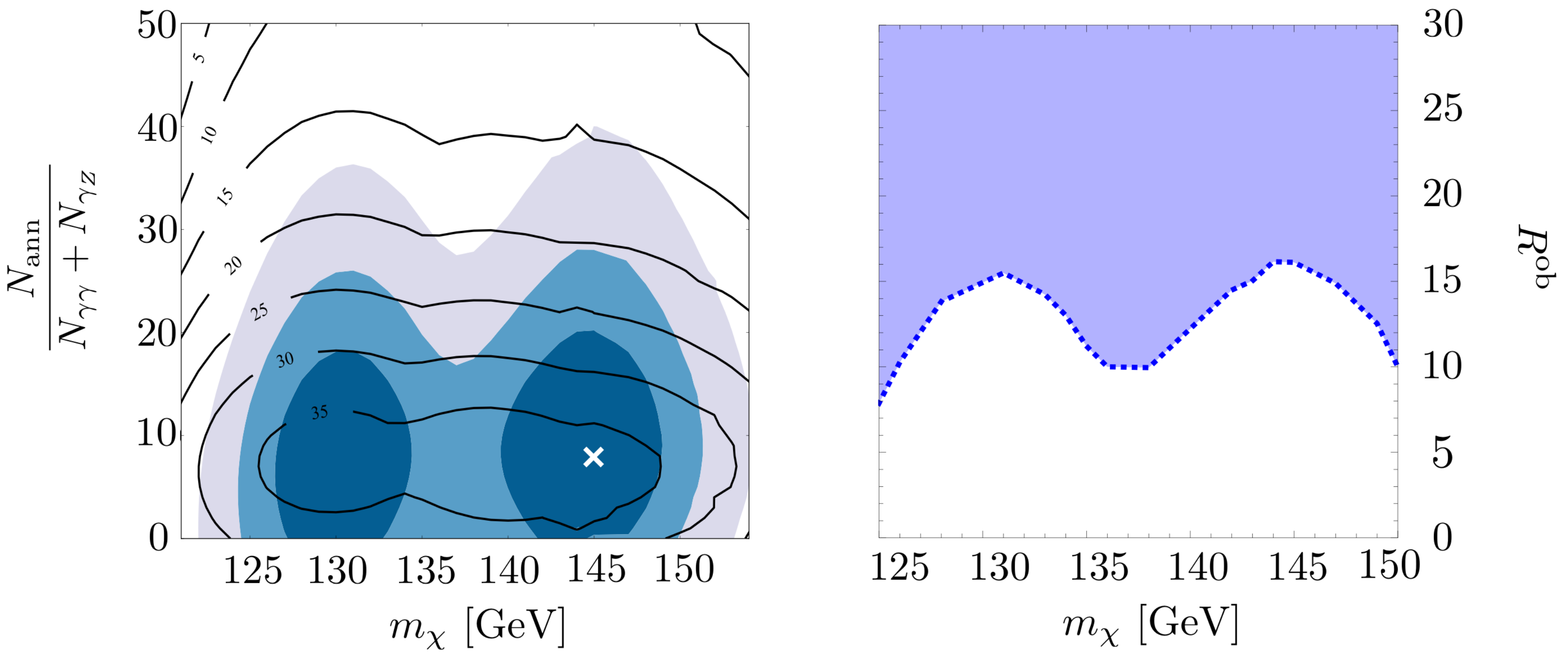}    
\includegraphics[width=.4\textwidth]{TemporaryDiagrams/colorbar.pdf} $\quad\quad\quad\quad\quad\quad\quad\quad\quad\quad\quad\quad\quad\quad\quad\quad\quad$
    \caption{The \emph{top} plot gives the photon counts within 3$^{\circ}$ degrees of the Galactic Center with the inner degree masked.  The solid red line shows the best fit model, which is given by the white cross in the bottom left plot.  This best fit point has  $\Nann \neq 0$; the dashed black line shows the continuum spectrum for 145 GeV dark matter annihilating into $ \tau^+ \tau^-$ with the best fit normalization.
On the \emph{bottom left}, we show 1, 2, and 3 $\sigma$ confidence regions (filled contours) for $\Nann/(\NGG+\NGZ)$ as a function of mass for dark matter annihilation to $ \tau^+ \tau^-$.  The ratio $\NGZ/\NGG$ is allowed to freely vary for each point in the grid.  The solid black lines are the contours for $\NGG+\NGZ$.  The best fit point is marked with a cross at $m_{\chi} = 145$ GeV, $\thetaGZ = 1.57$, and $\Nann = 284$.  On the \emph{bottom right}, we show the shape analysis constraint. The shaded region corresponds to parameters where the fit is 2 $\sigma$ or worse with respect to the best fit point.}
   \label{fig:ShapeConstraintTT}
\end{figure}

\newpage

\begin{figure}[h!!] %  figure placement: here, top, bottom, or page
   \centering
 {\Large $\chi \chi \rightarrow \mu^+ \mu^-$}\\
\vskip 10 pt
  \includegraphics[width=.9 \textwidth]{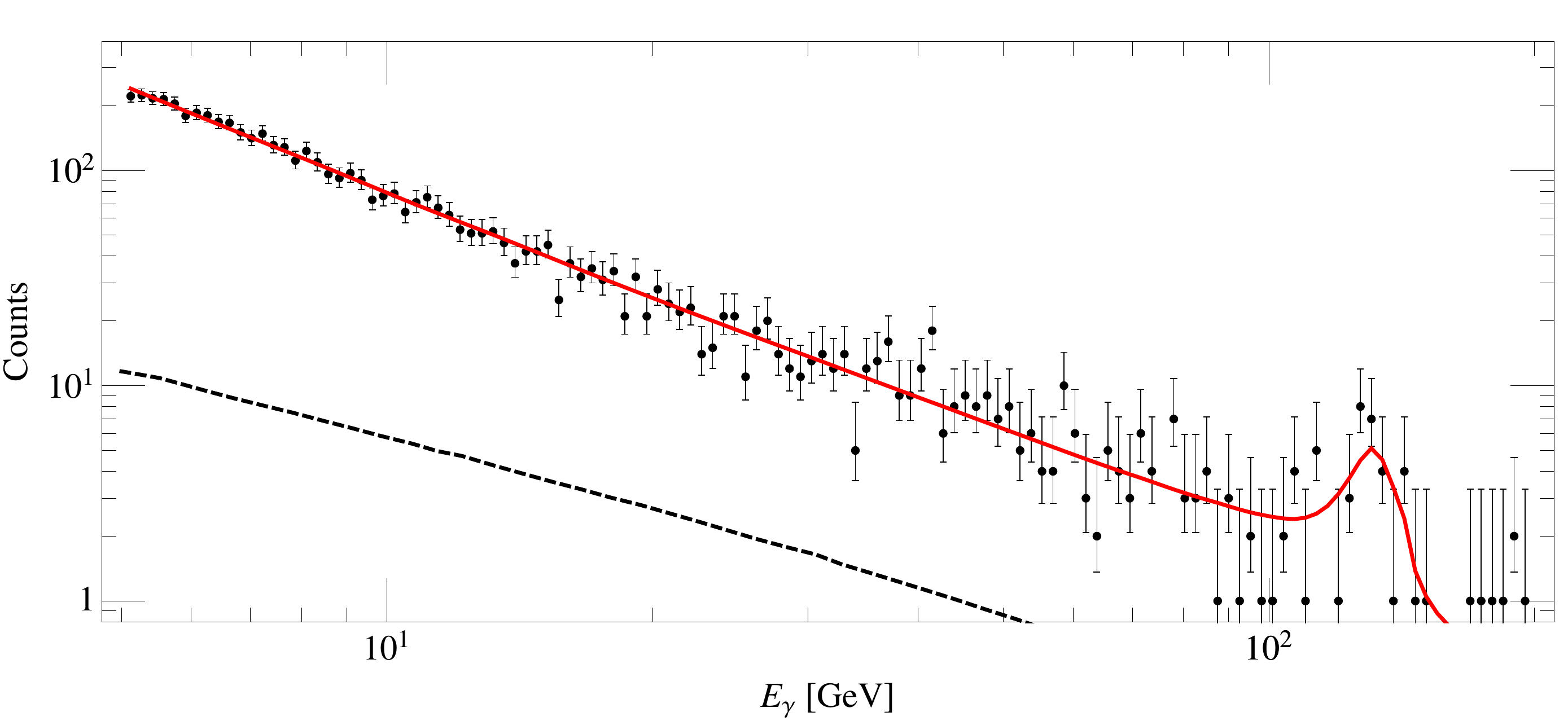}  \\
  \vskip 35 pt
\includegraphics[width=1.05\textwidth]{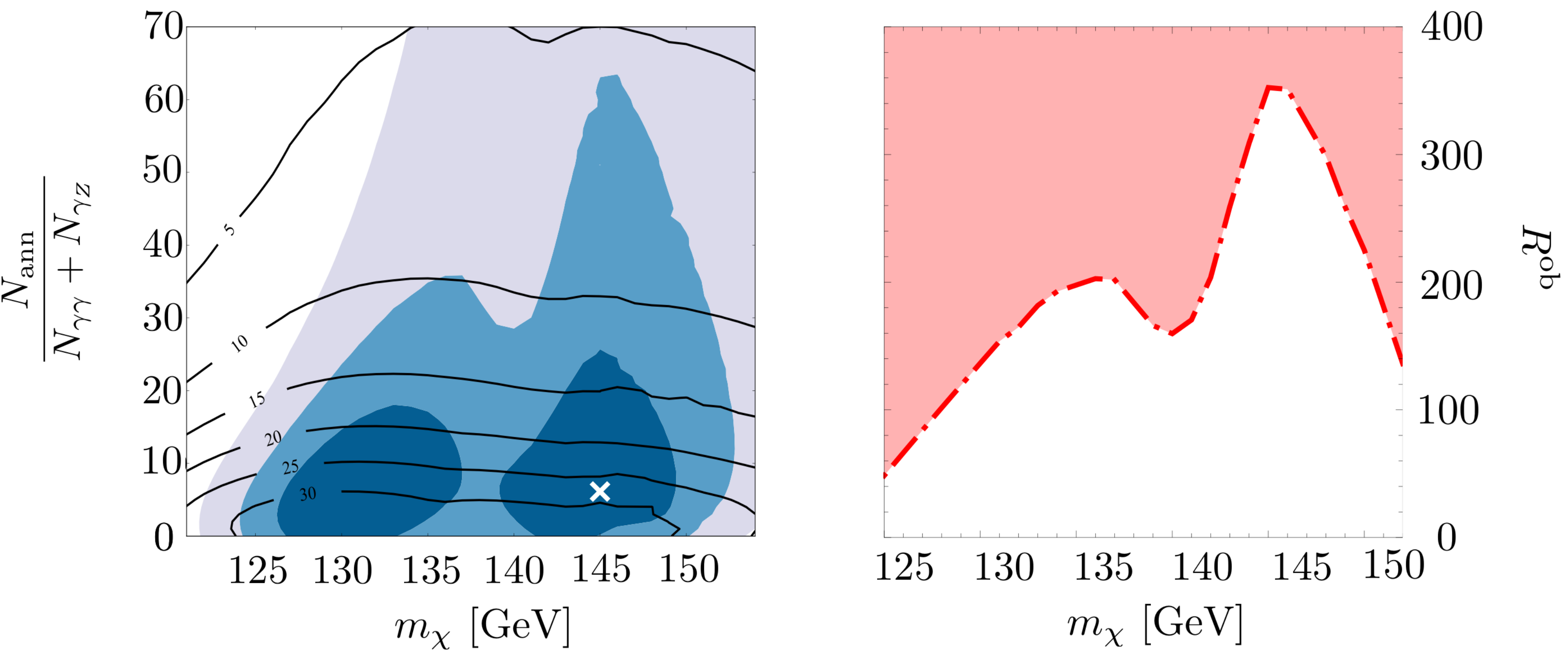}   
\includegraphics[width=.4\textwidth]{TemporaryDiagrams/colorbar.pdf} $\quad\quad\quad\quad\quad\quad\quad\quad\quad\quad\quad\quad\quad\quad\quad\quad\quad$
    \caption{The \emph{top} plot gives the photon counts within 3$^{\circ}$ degrees of the Galactic Center with the inner degree masked.  The solid red line shows the best fit model, which is given by the white cross in the bottom left plot.  This best fit point has  $\Nann \neq 0$; the dashed black line shows the continuum spectrum for 145 GeV dark matter annihilating into $ \mu^+ \mu^-$ with the best fit normalization.
On the \emph{bottom left}, we show 1, 2, and 3 $\sigma$ confidence regions (filled contours) for $\Nann/(\NGG+\NGZ)$ as a function of mass for dark matter annihilation to $\mu^+ \mu^-$.  The ratio $\NGZ/\NGG$ is allowed to freely vary for each point in the grid.  The solid black lines are the contours for $\NGG+\NGZ$.  The best fit point is marked with a cross at $m_{\chi} = 145$ GeV, $\thetaGZ = 1.57$, and $\Nann = 160$.  On the \emph{bottom right}, we show the shape analysis constraint. The shaded region corresponds to the parameters where the fit is 2 $\sigma$ or worse with respect to the best fit point.}
   \label{fig:ShapeConstraintMM}
\end{figure}

\pagebreak

\bibliography{GammaLineBibFile}
\bibliographystyle{utphys}

\end{document}